\documentclass{article}
\usepackage{graphicx}
\usepackage{amsmath}
\usepackage[super,comma,sort&compress]{natbib}
\usepackage{amssymb}
\usepackage{booktabs}
\usepackage{authblk}
\usepackage{url}
\usepackage{subcaption}
\usepackage{hyperref}
\usepackage{fullpage}
\usepackage{siunitx}
\usepackage{coffeestains}
\usepackage{tablefootnote}

\hypersetup{
    colorlinks=true,
    linkcolor=blue,
    filecolor=magenta,      
    urlcolor=cyan,
    pdftitle={Galaxy-Jet Alignment},
    pdfpagemode=FullScreen,
    }

\newcommand{\subsubsubsection}[1]{\paragraph{#1}}

\newcommand{\araa}{Annu. Rev. Astron. Astrophys.}   % Annual Review of Astron and Astrophys
\newcommand{\aj}{Astron. J.}   % Astronomical Journal
\newcommand{\ac}{Astron. Comput.} % Astronomy and Computing
\newcommand{\apj}{Astrophys. J.}   % Astrophysical Journal
\newcommand{\apjl}{Astrophys. J. Lett.}   % Astrophysical Journal, Letters
\newcommand{\apjs}{Astrophys. J. Suppl. Ser.}   % Astrophysical Journal, Supplement
\newcommand{\aap}{Astron. Astrophys.}   % Astronomy and Astrophysics
\newcommand{\cse}{Comput. Sci. Eng.} %Computing in Science and Engineering
\newcommand{\jg}{J. Geod.} %Journal of Geodesy
\newcommand{\mnras}{Mon. Not. R. Astron. Soc.}   % Monthly Notices of the RAS
\newcommand{\nat}{Nature} % Nature
\newcommand{\nastro}{Nat. Astron.} % Nature Astronomy
\newcommand{\pasa}{Publ. Astron. Soc. Aust.}   % Publications of the Astron. Soc. of Australia
\newcommand{\pasp}{Publ. Astron. Soc. Pac.}   % Publications of the Astron. Soc. of the Pacific
\newcommand{\sci}{Science} % Science
\title{Detection of an orthogonal alignment between parsec scale AGN jets and their host galaxies}

\author[1,2]{D.~Fernández Gil}
\author[1]{J.~A.~Hodgson}
\author[1]{B.~L'Huillier}
\author[3,4]{J.~Asorey}
\author[5,6,7]{C.~Saulder}
\author[8]{K.~Finner}
\author[9]{M.~J.~Jee}
\author[7,10]{D.~Parkinson}
\author[11]{F.~Combes}

\affil[1]{Department of Physics and Astronomy, Sejong University, 209 Neungdong-ro, Gwangjin-gu, 05006 Seoul, Korea}
\affil[2]{Centro de Estudios de Física del Cosmos de Aragón, P. San Juan, 1, 44001 Teruel, Spain}
\affil[3]{Departamento de Física Teórica and
Instituto de Física de Partículas y del Cosmos (IPARCOS-UCM), Universidad Complutense de Madrid, 28040 Madrid, Spain}
\affil[4]{Departamento de F\'isica Te\'orica, Centro de Astropart\'iculas y F\'isica de Altas Energ\'ias, Universidad de Zaragoza, 50009 Zaragoza, Spain}
\affil[5]{Max Planck Institute for Extraterrestrial Physics, Gie\ss enbachstra\ss e 1, 85748 Garching, Germany}
\affil[6]{Universit\"ats-Sternwarte M\"unchen, Scheinerstra\ss e 1, 81679 Munich, Germany }
\affil[7]{Korea Astronomy and Space Science Institute, 776 Daedeok-daero, Yuseong-gu, 34055 Daejeon, Korea}
\affil[8]{IPAC, California Institute of Technology, 1200 E California Blvd., Pasadena, CA 91125, USA}
\affil[9]{Yonsei University, Seodaemun-gu, Yonsei-ro 50, Seoul, Korea}
\affil[10]{School of Mathematics and Physics, University of Queensland, St Lucia 4072, Brisbane, Australia}
\affil[11]{LERMA, Observatoire de Paris, PSL Univ., Coll\`ege de France, CNRS, Sorbonne Univ., Paris, France}

\begin{document}

\maketitle

\begin{abstract}
    The relationship between galaxies and their supermassive black holes (SMBHs) is an area of active research. One way to investigate this is to compare parsec-scale jets formed by SMBHs with the projected shape of their kiloparsec-scale host galaxies.
    We analyse Very Long Baseline Interferometry (VLBI) images of Active Galactic Nuclei (AGN) and optical images of their host galaxies. We compare the inner-jet position angle in VLBI-detected radio sources with the optical shapes of galaxies as measured by several large optical surveys. In total 6273 galaxy-AGN pairs were found. We carefully account for the systematics of the cross-matched sources and find that Dark Energy Spectroscopic Instrument Legacy Imaging Surveys data (DESI LS) is significantly less affected by them. Using DESI LS, with which 5853 galaxy-AGN pairs were cross-matched, we find a weak but significant alignment signal (with a p-value $\lesssim$ 0.01) between the parsec-scale AGN jet and the kpc projected minor axis of the optical host galaxy in sources with measured spectroscopic redshifts. Our results show that the observed source properties are connected over 3 orders of magnitude in scale. This points towards an intimate connection between the SMBH, their host galaxies and their subsequent evolution.
\end{abstract}

One might naively expect the orientation of the Supermassive Black Hole (SMBH)-accretion disk system to be related to the optical orientation of its host galaxy. The answer to this question (one way or another) touches on many broad areas in astronomy, such as galaxy evolution, Active Galactic Nuclei (AGN) feedback and cosmology. Very Long Baseline Interferometry (VLBI) offers an unique opportunity to probe the pc to sub-pc scale jets \citep{2022ApJS..260...12W,2023MNRAS.520.6053P}, which may be a proxy for the orientation of the SMBH accretion disk system \citep{1995PASP..107..803U,2013Sci...339...49M,2019ARA&A..57..467B}. \\

This question has been explored by \citet{1981ApJ...246...28J} with 14 galaxy-VLBI jet pairs and inconclusive results. More recently this has been looked at by \citet{Mandarakas} using parsec-scale VLBI and \citet{Najar2019RadioOpticalAO} using kpc-scale radio jet position angles (PAs). A strong correlation between the radio jet and the optical minor axis of the host galaxies was not found, however these two results remain unpublished. Other authors have also searched for this correlation at lower resolutions with mixed results. \citet{1971MNRAS.151..421M} measured the kpc-scale major axis of both the radio and optical components of elliptical galaxies, finding a strong parallel correlation between them, while \citet{1979ApJ...231L...7P} used the same optical data but found an opposite, perpendicular correlation using different radio data. \\

\begin{figure*} \centering
    \includegraphics[width=0.8\textwidth]{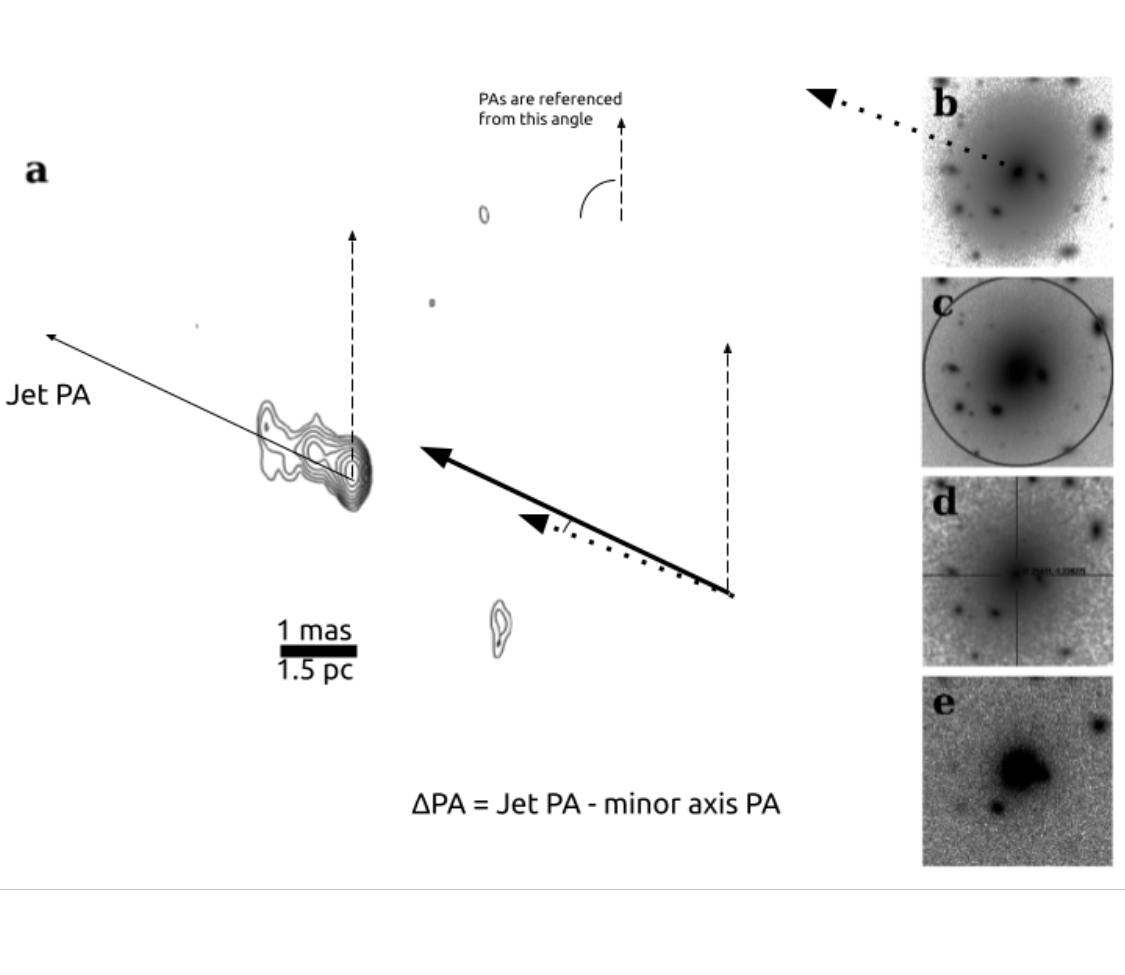}
    \caption{Demonstration of how the angle difference between the VLBI jet PA and the projected minor axis PA of an optical host galaxy is computed. a) VLBI image of J0029-0113 ($z=0.082817$) at observing frequency 8.7 GHz with an angular size of $50\times 50$ mas$^2$. b) Optical counter-part image from DES. c) Optical counter-part image from DESI LS. d) Optical counter-part image from SDSS. e) Optical counter-part image from SkyMapper. All optical images have a size of $1\times 1$ arcmin$^2$. One mas $\sim$ 1.5 pc ($H_{0} = 70 \rm{ km \, \rm s^{-1} Mpc^{-1}}$, $\Omega_\mathrm{m} = 0.3$). The PA angles are referenced North to East. We take the difference between the VLBI jet PA and the PA of the closest minor axis of the projected optical host galaxy shape.} \label{fig:VLBI_optical_images}
\end{figure*}

Other studies, looking at different but related topics, found no alignment evidence: \citet{2000ApJ...537..152K} computed the 3-D angle between the radio jet and disk for Seyfert galaxies; \citet{2002ApJ...575..150S} studied the three-dimensional orientation of jets relative to nuclear dust disks in 20 radio galaxies and \citet{2009ApJ...707..787G} looked at accretion disk maser candidates, which allowed analysis of the (mis)alignment between AGNs and their surrounding galactic stellar disks. \\

Some works used different approaches: \citet{2011MNRAS.414.2148L} simulated an AGN sample from non-AGN galaxies and found evidence that the angular momentum of material falling onto a BH influences the AGN. Of particular relevance to our study is the work of \citet{2009MNRAS.399.1888B} who compared the kpc-scale 20 cm radio and optical orientations of galaxies. The optical and radio orientations were found to be perpendicular to each other in radio-quiet elliptical galaxies. In comparison, VLBI detected sources are almost exclusively radio loud but also hosted in elliptical galaxies\citep{2001A&A...375..791V,2006AJ....132..531K,2012A&A...541A..62J,2022ApJ...941...95W}. \\

However, no previous study has approached this question by probing orders of magnitude closer to the SMBH-accretion disk system with VLBI while considering multiple optical surveys and carefully taking systematic errors into account. \\

The VLBI data is obtained from the Astrogeo VLBI Image Database which we use as the basis for cross-matching. Cross-matched optical data is found in the Sloan Digital Sky Survey Data Release 17 (SDSS), the Dark Energy Spectroscopic Instrument Legacy Imaging Surveys Data Release 10 (DESI LS), the SkyMapper Southern Sky Survey Data Release 2 (SkyMapper), the Kilo-Degree Survey 1000 Gold Sample (KiDS) and the Dark Energy Survey Data Release 2 (DES). For more details of the samples and surveys see the Methods Section. In Fig. \ref{fig:VLBI_optical_images}, our basic method is described. We take the minor axis of the projected shape of the host galaxy and the position angle (PA) of the jet. The key observable is the difference between these angles ($\Delta$PA) which lies between 0 and 90$^\circ$. 
Care needs to be taken in transforming the PA directions into $\Delta\text{PA}  \in [0,90^\circ]$ (see Section PA comparisons).

The footprints of the surveys are shown in Fig. \ref{fig:footprint}. The VLBI jet PA distribution is shown in Supplementary Figure \ref{fig:jetpa}, and two examples of the optical major axis PA distribution are shown in Supplementary Figure \ref{fig:opticalpa}: DESI LS in panel (a) and SkyMapper in panel (b). \\

 \begin{figure*} \centering
    \includegraphics[width=\textwidth]{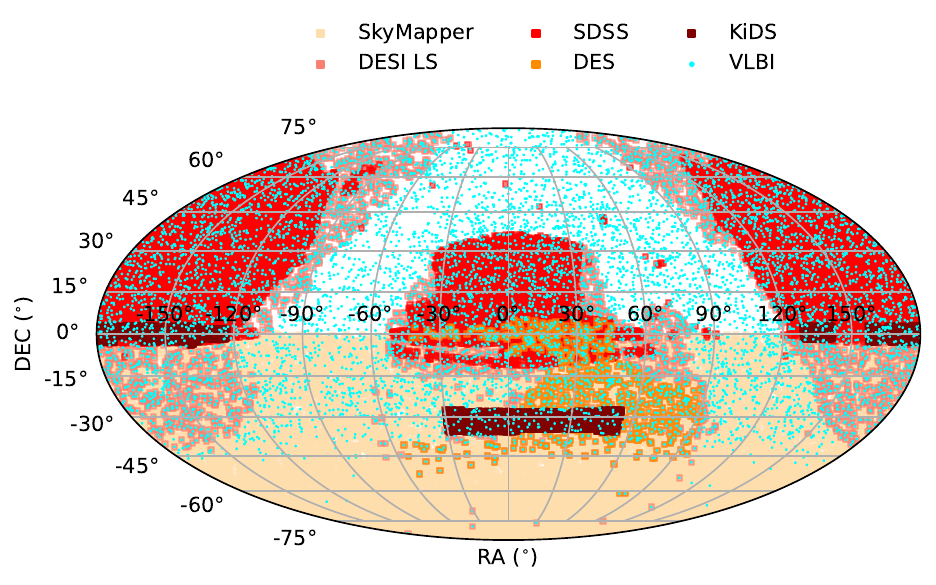}
    \caption{Footprints of the optical surveys and the VLBI sample used. Each survey is represented by a different color as shown in the legend.}
    \label{fig:footprint}
\end{figure*}

We take particular care to quantify the systematics (see Methods Section for a detailed description). It is unclear what is causing the systematic errors but nevertheless, we find that the DESI LS data is the least affected. For this reason, we proceed with the DESI LS data for the main conclusions. We show the result for the remaining individual optical surveys as well as a combined dataset in Supplementary Figures \ref{fig:combined_histogram}-\ref{fig:hist_SM}. \\

\begin{table*}[h!]
\centering
\caption{Information summary about the optical surveys. From left to right: name of the optical survey, median value of the Seeing (with the band it was measured in in parenthesis), total number of cross-matched sources, number of sources within the semi-minor axis cut (`criterion 0'), criteria for the `good case' (see the Optical surveys section in the Methods for more information), number of sources withing the semi-minor axis cut and the `good case', number of sources withing the semi-minor axis cut and the PA error cut.}\label{tab:listsurveys}
{%\tablefont
\begin{tabular}{@{\extracolsep{\fill}}lllllll}
\toprule
Survey  & Median   & Total cross- & $b >$ 1.3'' & Good case criteria & $b >$ 1.3''  & $b >$ 1.3''   \\ 
Name &  Seeing  & matches & (`criterion 0') &  & \& & \& \\
 &  &  &  &  & Good cases & $\sigma_{\text{PA}} < 22.5^{\circ}$ \\

\midrule

SDSS & 1.3" (r) & 2110 & 177 & type = \texttt{GALAXY} & 177 & 176  \\

DESI LS & 1.3" & 5853 & 534 & type = \texttt{SER}, \texttt{EXP}  or \texttt{DEV} & 524 & 520 
\\

KiDS & 0.7-0.9" (g) & 10 & & $\texttt{B\_IMAGE} >$ 2" & & 
\\
 &   &   & & \& $\texttt{CLASS\_STAR}<0.5$ & & 
\\

SkyMapper & $~$2.6" (g) & 1337 & 120 & $\texttt{b} >$ 2" \& $\texttt{CLASS\_STAR}<0.5$ & 56 & 36
\\

DES & 1.11'' (g) & 748 & 91 & $\texttt{EXTENDED\_CLASS\_COADD}$  & 83 & 26
\\
 &   &   &   &  = 2,3 &   &  
\\
\bottomrule
\end{tabular}}
\caption*{Text in \texttt{teletype} font denotes database parameters.}
\end{table*}
 
To compare between VLBI jet PAs and optical PAs it is necessary to filter out unresolved sources. To do this, we first remove any cross-matched source with a semi-minor axis $<$ 1.3 arcsec (`criterion 0') due to the mean seeing limits of DESI LS (see Table \ref{tab:listsurveys}). We then perform several complementary analyses. In the first case, we cut out all sources where the error in either the optical PA ($\sigma_\mathrm{PA,opt}$) or the VLBI PA ($\sigma_\mathrm{PA,VLBI}$) is higher than $\pm22.5^{\circ}$. Any PA measurements with errors larger than this are too unreliable for our analysis. 
In the second case, we use various criteria for each survey (as described in the Methods section) to determine if an optical PA belongs to a `good case'. Not all surveys contain the same morphological and resolution information. The criteria to consider galaxies a ‘good case’ need to be adapted for each individual survey, but we attempt to make them as consistent as possible. Even though the `good case' cut could be not as homogeneous as desired, as we proceed only with the DESI LS cross-matched sources for the main analysis, the impact of the sample selection diversity is negligible. We are motivated to perform the `good case' and minor axis cuts to try to differentiate between sources that are resolved but mostly circular from sources that are simply unresolved. We refer to the previous three cuts as `quality cuts'. We can also speculate that any correlation may have a redshift dependency, as closer sources would be better resolved on average. For this reason, we also make cuts with a) all sources with spectroscopic redshifts, b) sources with spectroscopic redshifts below 0.5 and c) sources with spectroscopic redshifts below 0.1. \\

 \begin{figure*} 
    \includegraphics[width=\textwidth]{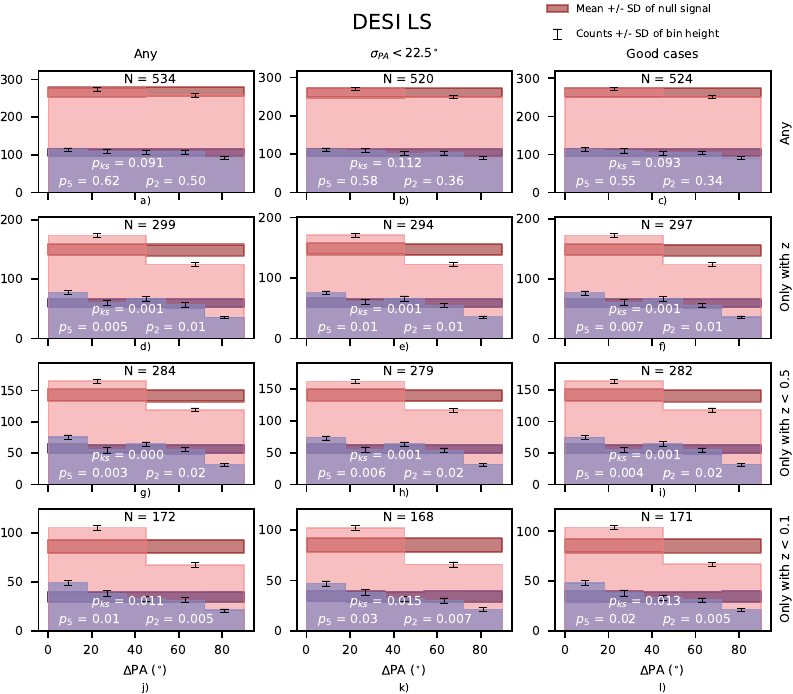}
    \caption{Histograms of all VLBI sources with a cross-match in DESI LS and a semi-minor axis $>$ 1.3" (the `criterion 0'). $\Delta\text{PA}$ indicates the angle between the VLBI jet and the optical minor axis of its host galaxy. The dark bands show the expected null signal (no alignment), which is the result of our careful analysis of the systematic errors. The stronger the deviation from the dark band the stronger the alignment signal. $p_5$ and $p_2$ are the p-values for five bins and two bins in the $\Delta\text{PA}$ space respectively. $p_{\text{ks}}$ is the p-value from a KS test comparing the data against a uniform distribution, independently of the number of bins. Each panel corresponds to a specific selection of sources, showing the number of sources on top. Each panel corresponds to a specific selection of sources based on several cuts. The number of sources inside each panel is shown on top. a) All `criterion 0' sources. b) All `criterion 0' sources whose jet and optical PA error is $<$ $22.5^{\circ}$. c) All `criterion 0' sources within the `good case'. d) All `criterion 0' sources with a measured spectroscopic redshift. e) All `criterion 0' sources whose jet and optical PA error is $<$ $22.5^{\circ}$ and with a measured spectroscopic redshift. f) All `criterion 0' sources within the `good case' and with a measured spectroscopic redshift. g) All `criterion 0' sources with $z < 0.5$. h) All `criterion 0' sources whose jet and optical PA error is $<$ $22.5^{\circ}$ and with $z < 0.5$. i) All `criterion 0' sources within the `good case' and with $z < 0.5$. j) All `criterion 0' sources with $z < 0.1$ k) All `criterion 0' sources whose jet and optical PA error is $<$ $22.5^{\circ}$ and with $z < 0.1$. l) All `criterion 0' sources within the `good case' and with $z < 0.1$.}
    \label{fig:hist_desi}
\end{figure*}

In Fig. \ref{fig:hist_desi} we show histograms of the $\Delta$PA distribution for two and five bins of the DESI LS cross-matches for the different cuts that were described in the previous paragraph. The left column has no `quality cuts', the center column contains only sources with a VLBI or optical PA error $\sigma_\mathrm{PA}$ below the cut and the right column contains only the `good case' sources. Additionally, each row applies a cut in redshift: the top row has no cuts, the second row only contains sources that have a spectroscopic redshift and the two lower rows only contain sources with a spectroscopic redshift below 0.5 and 0.1 respectively. Each subfigure contains two histograms, one with five bins and one with two bins. Each histogram includes a dark band which represents the expected null signal obtained by reshuffling the sample of the jet PA 1000 times. The band is centered around the average height of each bin after the reshufflings, and the (vertical) width is the standard deviation. The histogram shows the distribution of $\Delta$PA and the error bars of the bins are obtained from Monte Carlo simulations. The larger the deviation from the dark band, the stronger the alignment signal (i.e. the p-value is lower).

The 2-bin histogram is justified since it tells us immediately if there are more pairs with $\Delta\text{PA} < 22.5^\circ$, i.e., if we have a preference for an alignment or an anti-alignment. We investigated the effect of binning by generating results with 4,5,6,8, and 10 bins. We show the 10-bin histogram Supplementary Figure \ref{fig:DESI_10_bins}. Changing the binning sometimes increases and sometimes decreases the p-value, but does not change the overall picture. We ultimately choose to include the 5-bin histogram in the main Figure (Fig. \ref{fig:hist_desi}) as it allows for a finer analysis while keeping a statistically significant number of pairs per bin. We additionally perform a Kolmogórov-Smirnov (KS) test and include the p-values, as it does not depend on binning. The KS test is a non-parametric test that determines the goodness of fit between two different distributions. The standard test, which is the one we use, takes a given set of data and compares it against a uniform distribution.

We further investigated the effect of magnitudes. In general cutting by magnitude reduces the number of sources, especially in the top two rows. We demonstrate this in the Methods section, where we apply a strict magnitude cut in z-band $< 14$ (shown in Supplementary Figure \ref{fig:DESI_mag_cuts}) and compare the histogram with Fig. \ref{fig:hist_desi}. As before, applying these cuts affects the number of sources but the results remain consistent.\\

Our results show a weak but significant alignment ($p \leq 0.03$) in both the two and five-bin cases for both `quality cut' sources and 'any' sources with a measured spectroscopic redshift. No significant signal is seen when including sources without spectroscopic redshifts. The significance is highest $p \leq 0.007$ for all sources with measured redshifts and for sources with $z < 0.5$. \\

In the `good cases' cut with redshift $z < 1$ we find a median $\Delta$PA of ${40^{+34}_{-27}}^{\circ}$ with a 68\% error. We also investigated if the median evolves with redshift in Supplementary Figure \ref{fig:DESI_delta_redshift} but found no clear trend. Overall, we find good evidence for a weak alignment between the pc-scale AGN jet PA and the kpc-scale minor axis of the projected shape of the host galaxies for sources with a well-measured optical shape and spectroscopic redshift. \\

We also visualize the correlation between both angles with a scatter plot in Supplementary Figure \ref{fig:DESI_scatterplots}, but this does not provide a clear visible result due to the weak nature of the alignment. To help visualization we use a color map and add a histogram with the counts per bin to the $\Delta$PA scatter plot in Supplementary Figure \ref{fig:DESI_scatter_hist}. \\

The original assertion was that we might expect a connection between the SMBH and the general properties of the galaxy. 
This is particularly true in VLBI sources that are mostly hosted in radio-loud elliptical galaxies \citep{2001A&A...375..791V,2012A&A...541A..62J,2022ApJ...941...95W}. In spiral galaxies, which are mostly star-forming galaxies, the radio emission is often due to synchrotron emission from supernovae, which traces the star-forming disk. This could explain why \citet{2009MNRAS.399.1888B} reported a correlation between the optical major axis of the galaxy and the major axis of the radio images, as taken from the FIRST-survey.  Even in the case of nearby jetted spirals, an alignment of the jet with the large-scale optical disk is not expected \citep{2000ApJ...537..152K}.  \\

A critical component to interpreting the result is the physical meaning of the projected minor axis of the host galaxy. Projection effects could be complex, but they cannot create such an alignment alone. On one hand, simulations suggest that there could be dynamical reasons why radio-loud elliptical galaxies (E-gals) would show a better alignment between their SMBH accretion disk systems and their large-scale optical morphology, in comparison to spiral galaxies. Spiral galaxies are mainly formed through secular evolution and gas accretion from cosmic filaments, while E-gals are often formed by major mergers \citep {2013ARA&A..51..511K}. These different formation scenarios have alignment consequences with cosmic filaments \citep{2018MNRAS.481.4753C}. Simulations further suggest that the remnant of a major merger would be a spheroidal object elongated along the arrival direction of the swallowed companion, with a small disk forming from the tidal debris, slowly rotating around one of the minor axes of the remnant. The gas from this disk will be accelerated from the SMBH while conserving its angular momentum, resulting in a jet that is perpendicular to the accretion disk. In spiral galaxies, both cosmic gas accretion and minor mergers occur isotropically, which results in a final random accretion direction onto the SMBH. Small radio jets in spirals often encounter disk material and are bent several times, so that the outflow eventually escapes from the disk material in a perpendicular direction at large scales. \\

\begin{figure*} 
    \includegraphics[width=\textwidth]{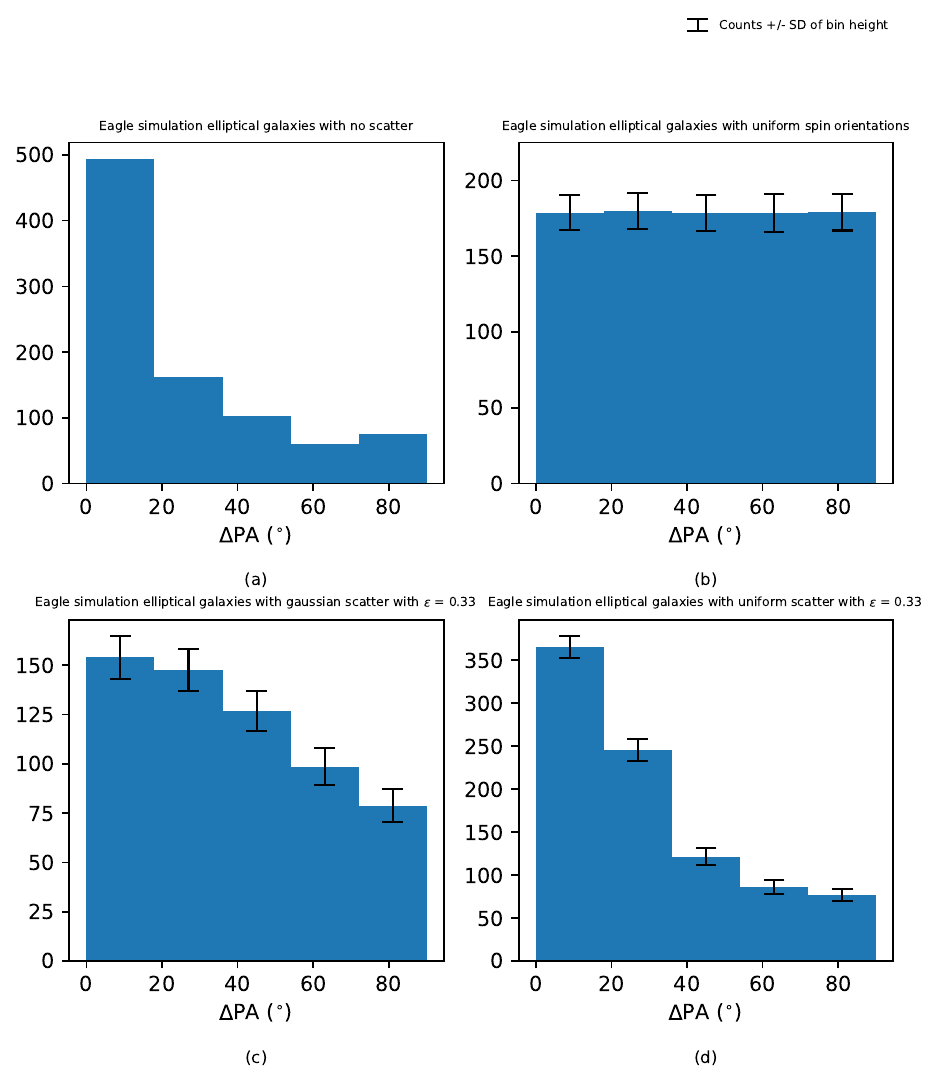}
    \caption{$\Delta$PA between the projected spin axis of the stellar component and the semi-minor axis of the projected ellipse of the elliptical galaxies in the Eagle simulation. $\epsilon$ indicates the amplitude fraction of the offset we introduce to the 3D spins. The histogram is the result of averaging 1000 simulations, each time sampling the spin tilt offset errors randomly from different distributions as explained in the text above and the Methods Section. The error bars are the standard deviations. a) No offset. b) Uniform spin orientations. c) Gaussian scatter with $\epsilon = 0.33$. d) Uniform scatter with $\epsilon = 0.33$.}
    \label{fig:Eagle_sim}
\end{figure*}

We can use the aforementioned simulations to better understand the physical meaning of the signal. We used the EAGLE simulation 
\cite{EAGLE_sim_1,EAGLE_sim_2} and explored 4 scenarios: i) no modifications to the simulation data with a close 3D relationship between the stellar component spin of the galaxy and the main plane of the galaxy (see Methods for more details), ii) randomizing the 3D relationship, iii) adding a gaussian and a iv) uniform scatter to the original close relationship of 33$^{\circ}$ as motivated by the 68\% errors on the median of the $\Delta$PA distribution in the DESI LS `good cases' with $z<1$. In the first scenario (panel (a) of Fig. \ref{fig:Eagle_sim}) 
we get a very strong alignment signal. In the second scenario (panel (b) of Fig. \ref{fig:Eagle_sim}), we recover a uniform signal as expected, and in the third scenario (panel (c) of Fig. \ref{fig:Eagle_sim}) we interestingly recover a qualitatively similar-looking distribution to our observed distributions in the DESI LS `quality cuts' with low redshift, with a similar median and 68\% error in $\Delta$PA. The fourth scenario (panel (d) of Fig. \ref{fig:Eagle_sim}) shows a slightly stronger signal.

While the simulations could be suggestive, we should be careful not to over-interpret as there are many reasons one can imagine that could weaken such an alignment. The direction of jets may depend on the radio observing frequency and can vary on time-scales of the order of AGN duty cycle, i.e. 10 Myr \citep{2016A&A...595A..46M}, much smaller than the time-scale of optical morphology variation for galaxies (several Gyr). The jet direction depends on the angular momentum of the accreted material, and gas can come from a gas-rich late minor merger. Similarly, because mergers are complex dynamical processes the angular momentum connection between the SMBH and the kpc-scale structure could be scrambled \citep[e.g.,][]{1991ApJ...370L..65B}. A possibility that may weaken the minor axis alignment and has been seen in some cases \citep[e.g.][]{2019ApJ...871..143P} is that the optical jet or dust dominates the morphology of the host galaxy, which explained the previously observed offset in VLBI and Gaia sky positions \citep{kovalev17,kovalev20}. In this case, the optical shape would correspond to the jet rather than that of the galaxy, and the inferred galaxy position angle would be the direction of the jet. This could yield an alignment with the major axis. In the case of dust emission, it can also effect positional offsets between VLBI and Gaia, which could also affect the alignment signal reported here. All else being equal, complicated effects would be expected to weaken any observed signal. 

Simulations also show that misalignments arise naturally from the angular momentum transfer and therefore only a weak connection would be predicted and even then only for maximally spinning BHs \citep{2012MNRAS.425.1121H}.\\

A connection between the morphologies over such large scales could support the hypothesis of co-evolution of the BH and the host galaxy \citep{2013ARA&A..51..511K}, however, the interpretation of the result in this context remains unclear.

If the parsec scale jets are indeed a proxy for the SMBH-accretion disk system then this points to connections over an even larger range of scales. Observationally, the link between the SMBH and the inner jet sometimes seems clear: \citet{gomez2016} probed the innermost jet of BL Lacertae with the \emph{RadioAstron} space VLBI antenna, \citet{boccardi2016} probed the two-sided jet in Cygnus A with the Global mm-VLBI Array (GMVA) and Event Horizon Telescope (EHT) observations of both M87 and Cent A \citep{EHT_main,lu_2023,janssen2021}.  However, other studies were not so clear: \citet{hodgson2017} used the GMVA and \citet{gomez2022} used \emph{RadioAstron} to find parsec scale jet bends in OJ 287, \citet{giovannini2018} found large cross jet opening angles in 3C 84 using \emph{RadioAstron}, which was confirmed using the GMVA by \citet{oh2022} and \citet{paraschos2022}. Similarly, \citet{jy2020} used the EHT to find sub-parsec scale jet bends in 3C 279. \citet{oh2022}. Perhaps crucially, however,  further downstream (parsec-scale) the jets seem to follow the overall jet direction. The connection of the SMBH to the accretion disk system itself is also unclear \citep[e.g.][]{combes2019} and will require further study. Nevertheless, that we see a correlation could suggest either that the momentum of the SMBH accretion disk system is influencing or being influenced by the host galaxy and/or that the correlation is more influenced by the interplay between the jet itself and the host galaxy (i.e. AGN feedback). \\

An important finding is that having a well-measured shape and spectroscopic redshift introduces the alignment signal. We see in the last row of Fig. \ref{fig:hist_desi} that the signal is most significant, but the improvement is not so large. This is largely to be expected as the number of sources in that row also reduces.  Also, the AGN density peaks around $z \sim 1-2$, and therefore there are not many radio-loud AGN at $z<0.1$. Another reason is that at low $z$ we observe smaller volumes. Having a measured redshift introducing the signal suggests a strong resolution effect as sources with $z>0.5$ rarely have well-measured shapes. This is reflected in the relatively small change in source counts between any redshifts and $z<0.5$. \\

Our examination of the `quality cuts' seems to indicate a scarcity of genuinely spherical galaxies associated with a radio-loud jet. This implies that nearly all radio-loud elliptical galaxies possess a non-spherical form, potentially linked to the SMBH and accretion disk system. If this linkage is also discovered in spiral galaxies and radio-quiet AGN, it could point towards a more universal correlation. However, it is difficult to probe the orientation of the SMBH-accretion disk system without jets.

\section*{Conclusion}
 We find good evidence that the pc-scale VLBI jet is connected to the projected kpc-scale optical host galaxy morphology. The direction of the pc-scale jet (which may be a proxy for the SMBH-accretion disk system) appears to be oriented perpendicularly to the projected optical shape of the host galaxy with a $p$-value $<$ 0.05. The interpretation of these results is complicated but may have far-reaching implications. Ultimately these results will need to be followed up with higher source counts and higher-resolution optical observations (e.g. HST, JWST)  and with radio-quiet VLBI observations in order to determine how universal such a connection may be. In the near future there will be large surveys such will be performed using the Vera C. Rubin Observatory and Square Kilometer Array (SKA) VLBI and therefore the potential to vastly increase our sample sizes.

\appendix

\section*{Methods}\label{sec:methods}

\renewcommand{\thefigure}{M\arabic{figure}}
\setcounter{figure}{0}

\subsection*{Observations}\label{sec:observations}

\subsubsection*{VLBI Jet images and Astrogeo catalogue}\label{sec:Astrogeo}

VLBI is a technique that allows for radio antennae, that can be separated by many thousands of kilometers (or even into space), to be combined to create ultra high angular resolution images. This angular resolution is set by the ratio $\lambda /D$ where $\lambda$ is the observing wavelength and $D$ is the distance between the most spatially separated antennae.  
VLBI has been used to image black hole shadows \citep{2019ApJ...875L...1E} and to measure cosmic distances \citep{2020MNRAS.495L..27H}.  \\

The Astrogeo VLBI image database contains over 112,000 images of over 17,000 sources which have been collected from various VLBI surveys \citep[e.g.][]{2009JGeod..83..859P,2012A&A...544A..34P}. The observations are performed at observing frequencies ranging from 1.4 to 86 GHz. Most observations are performed at 2 GHz, 5 GHz, 8 GHz, or 15 GHz. Because these observations use globally located antennae, angular resolutions of sub-mas are regularly achieved. These angular resolutions often correspond to sub-pc scales even at relatively high redshifts ($z>1$). In Fig. \ref{fig:VLBI_optical_images}
an example VLBI source at parsec scale is displayed. This scale allows us to probe the central BH-accretion disk system of the galaxy. In contrast, the optical images at the corners are taken from the various optical surveys (described in more detail in the next section) and show the host galaxy in vastly larger (kpc) scales. \\

Typically, the sources detected in the VLBI observations are the bright, compact cores of radio AGN jets. These sources frequently exhibit a core-jet structure where the core is the bright component most upstream in the jet and the jet emission is fainter, extended and emanates from the core.  \\
 
In cases where there is a clear core-jet structure, a jet position angle (PA), i.e., the direction of the jet emission relative to the core, can be reliably determined. This analysis was performed by \citet{Plavin22}, who found reliable PAs for 9220 sources. The jet PAs were measured by fitting two Gaussian components directly in the uv-plane using a Bayesian nested-sampling algorithm, which also provides formal errors. They then compared these formal errors against the intrinsic scatter over multiple epochs and/or frequencies in order to estimate the true errors. \\
 
 These sources form the basis of our VLBI dataset. Redshifts are obtained from the OCARS \citep[Optical Characteristics of Astrometric Radio Sources,][]{2018ApJS..239...20M} catalog. The catalog differentiates between spectroscopic, photometric and generally unreliable redshift measurements. When redshifts are used, we only use spectroscopic redshifts. These are found for roughly 75\% of sources with a redshift measurement, and 43\% of all sources.

\subsubsection* {Optical surveys}
\label{sec:opticals}
For the orientation of the host galaxies we are using photometric images from five optical surveys. These surveys are listed in Table \ref{tab:listsurveys}: Sloan Digital Sky Survey DR17 (SDSS), DESI Legacy Imaging Surveys DR10 South and DR9 North (DESI LS), Kilo-Degree Survey 1000 Gold Sample (KiDS), Skymapper Southern Sky Survey (SkyMapper) and the Dark Energy Survey (DES). In all cases optical shapes are obtained from catalogues provided by these surveys. Fig. \ref{fig:footprint}
shows the footprint of each survey, and details are given below.

\subsubsubsection{Sloan Digital Sky Survey DR17 (SDSS)}
\label{sec:SDSS}
The Sloan Digital Sky Survey (SDSS) is an optical survey with imaging and spectroscopic information covering roughly 14,500 square degrees of the sky in both hemispheres \citep{SDSS1,SDSS2}. We use the latest data release, DR17 \nocite{SDSSDR17}, which belongs to phase IV of the survey. SDSS used the Sloan Foundation 2.5m Telescope at Apache Point Observatory in New Mexico, USA. The imaging data used here was taken by SDSS’s imaging camera \citep{Gunn:1998} taking data in the u, g r, i, and z bands.  \\

We access the galaxy photometric information through the PhotoObjAll table, which provides, among others things, the magnitude, the PA of the major axis and the minor-major axis ratio in 5 different bands: $u,g,r,i,z$. SDSS fits two models to each object in each band: a pure de Vaucouleurs (DeV) profile and a pure exponential (exp) profile. Each model has an arbitrary axis ratio and position angle. The models are convolved using a double-Gaussian fit to the PSF and the likelihood associated with the $\chi^2$ fit is computed. Although for large objects it is possible and even desirable to fit more complicated models (e.g., bulge plus disk), the computational expense to compute them is not justified for the majority of the detected objects \citep{SDSSDR17}. They provide the PA (\texttt{Phi}) and axis ratio (\texttt{AB}) parameters for each model, plus the \texttt{modelMag} parameter, which is the magnitude corresponding to the better of \texttt{DeV}/\texttt{Exp} magnitude fits in the r band. For our purposes, we use the best of these two models only in the r band. So, we would access the PA of an specific source in the r band using the \texttt{deVPhi\_r} parameter if the best-fit model is DeV or \texttt{expPhi\_r} parameter if it is Exp. To make sure that the choice of band did not strongly affect the results, we compare the optical PA between all the different bands and see that they are mostly correlated, with some having an offset of $180^{\circ}$ due to the galaxy symmetry. The errors on the major axis PAs are not provided since DR8 because systematic errors dominate over the photon statistical errors. These systematic errors are of the order of 15 degrees \citep{url1}. We adopt this value for all SDSS sources analysed in this study. Finally, we also make use of the morphological classification \texttt{type} (3 for resolved or \texttt{GALAXY} , 6 for unresolved or \texttt{STAR}) of the object, which is computed comparing the magnitudes of the PSF and the fitted model.

\subsubsubsection{Skymapper Southern Sky Survey DR2 (SkyMapper)}
SkyMapper Southern Sky survey is a wide-field optical survey that covers over 21,000 square degrees using the SkyMapper 1.35 m telescope at Siding Springs Observatory in Australia. Using its own wide-angle 268 Megapixel camera, the survey focused on creating a full survey of the southern sky. In particular, the second data release (DR2) contains up to 500 million objects between AB magnitudes 8 and 22 in the u,v,g,r,i,z bands \citep{2018PASA...35...10W,2019PASA...36...33O}. We use the DR2 photometry table to obtain the shape parameters. They use Petrosian models to fit the parameters to the galaxies. Once the Petrosian radius is computed, its position angle and axis ratio are obtained. The major axis PA is provided in the \texttt{pa} parameter. Unlike in SDSS, the error information of the PAs is provided in the \texttt{e\_pa} parameter and included in our analysis. This catalogue does not provide a systematic classification of objects between point like and extended, but they provide \texttt{CLASS\_STAR}, the stellarity index from \texttt{SExtractor}; and the semi-minor axis \texttt{b}. \texttt{CLASS\_STAR} defines an $a$ $posteriori$ probability of a detection to be a point source or an extended object. It is usually closer to zero for extended sources and close to one for point-like sources \citep{url2}.

\subsubsubsection{DESI Legacy Imaging Surveys DR10 (DESI LS)}\label{sec:DESI}
The DESI Legacy Imaging Surveys \citep{2019AJ....157..168D} consist of three separate projects: the
Dark Energy Camera Legacy Survey (DECaLS), the Beijing-Arizona Sky Survey (BASS), and the Mayall z-band Legacy Survey (MzLS). The data from BASS and MzLS, which covers a large part of the North galactic cap North of a declination of 32.375$^{\circ}$ is the same as from the previous iteration of the DESI Legacy Imaging Surveys DR9. The current data release DR10 has been supplemented by new data observed with the Dark Energy Camera, mainly on the Southern hemisphere. This includes data from the Dark Energy Survey, the DELVE Survey, and the DeROSITA Survey. The surveys conducted with the Dark Energy Camera in the g, r, and z-band were carried out on Blanco 4m telescope, located at the Cerro Tololo Inter-American Observatory in Chile. The BASS was carried out using the 90Prime camera with g and r band filters at the Bok 2.3-m telescope, while the MOSAIC-3 camera with a z band filter at the 4-meter Mayall telescope was used for the MzLS. Both the Bok and the Mayall telescope are located next to each other at Kitt Peak National Observatory in Arizona, USA. \\

On this wealth of data, the {\sc Tractor} \citep{url3} pipeline is used to obtain photometric models for all detected sources across multiple imaging bands. These galaxy model parameters were obtained directly from {\sc Tractor} (Lang et al, in preparation), which fits simple elliptical models consisting of single elliptical isophotes and suitable radial brightness profiles for the various morphological types (\texttt{REX}, \texttt{EXP}, \texttt{DEV}, and  \texttt{SER}) to the DESI Legacy Imaging Surveys. Using these models, the parameters are measured by optimizing a pixelwise forward model, correlating each model image with the individual image PSFs.Using the {\sc sweep} catalogues created by {\sc Tractor}, we obtained fluxes and shape parameters like the PA and its errors (\url{}). As the shape measurements are obtained jointly across all available optical bands, we select only objects from which the catalogue has non-zero fluxes in the g, r, and z band with finite measurement uncertainties.  \\

To each of the detected sources, {\sc Tractor} fits a range of morphological models: \texttt{PSF} (stellar, unresolved), \texttt{REX} (round exponential galaxy with a variable radius), \texttt{EXP} (exponential), \texttt{DEV} (de Vaucouleurs), \texttt{SER} (Sersic), \texttt{DUP} (sources detected by the Gaia Space Telescope but not by DESI LS). The morphological type used for the set of parameters provided for each source is determined by the $T$ value of the model, which also gets penalized (\url{https://www.legacysurvey.org/dr9/description/#morphological-classification}) for more ``advance models'', with \texttt{PSF} being the simplest, then \texttt{REX}, followed by \texttt{EXP} and \texttt{DEV} and finally \texttt{SER} as the ``best'' model. The \texttt{DEV}, \texttt{EXP}, and \texttt{SER} models provide shape fits for extended sources and with them: S\'ersic indices $n$,  half-light radii $r_\mathrm{eff}$, the complex ellipticities ($\epsilon_1+i\epsilon_2$) from where the axis ratio $\frac{b}{a}$, axis PA $\phi$, and their errors can be obtained using the following equations:  \\

\begin{align}
    \bar{\epsilon} & = \sqrt{\epsilon_{1}^{2}+\epsilon_{2}^{2}}, \nonumber\\
    \frac{b}{a} &  = \frac{1-\bar{\epsilon}}{1+\bar{\epsilon}},\\
    \phi & =\frac{1}{2} \textrm{arctan}\left(\frac{\epsilon_{2}}{\epsilon_{1}}\right).
\end{align}
\\

\subsubsubsection{Kilo-Degree Survey 1000 Gold Sample (KiDS)}\label{sec:KIDS}
The Kilo-Degree Survey (KiDS) is an optical wide-field imaging survey designed to measure weak gravitational lensing using OmegaCAM on ESO's VLT Survey Telescope at Cerro Paranal in Chile. It has a sky coverage of 1,500 square degrees with imagaing data in five near infrared bands: Z, Y, J, H, and K. and 4 optical bands: u, g, r and i.. The KiDS-1000 data set only contains galaxies with reliable shape and redshift measurements, their ``gold sample''. They provide photometric shape parameters by using their Astro-WISE image processing pipeline, which creates co-added images combining all bands, and then applying \texttt{SourceExtractor} to them. \texttt{SourceExtractor} works doing isophotal measurements to the filtered, background-subtracted detection images. It only uses pixels with values above the detection threshold and fits a single ellipse into them, with the corresponding position angle and axis ratio. This is a very quick process, but it is very dependent on the detection threshold itself. As mentioned before, more complex models could be fitted to some of our galaxies, but this is outside the scope of the paper, which aims to focus on a first detection of correlations. The major axis PA and its error are provided in the \texttt{THETA\_J2000} and \texttt{ERRTHETA\_J2000} parameters, and we use the stellarity index \texttt{CLASS\_STAR} (another parameter provided by \texttt{SourceExtractor}) and the minor axis \texttt{B\_IMAGE} to determine the resolution of the sources like in SkyMapper (see the Section on SkyMapper; \citealt{2013Msngr.154...44D}). Only ten VLBI sources are found to cross-match with the KiDS survey data.

\subsubsubsection{Dark Energy Survey DR2 (DES)}\label{sec:DES}

The Dark Energy Survey (DES) is a ground-based, wide-area, optical and near-IR imaging survey that took place from 2013 until the last observing season in 2019. For this, the DES collaboration designed, built and used the Dark Energy Camera (DECam), a 570 Mpix camera at the 4m Blanco telescope in Chile \citep{2008arXiv0810.3600H,2015AJ....150..150F}. For their cosmology goals, DES implemented two surveys: a main wide survey covering about $5000 \deg^2$ in the grizY photometric bands and a $30 \deg^2$ deep supernova survey in the griz bands. The data is processed at the National Center for Supercomputing Applications (NCSA) using the DES pipeline \citep{2018PASP..130g4501M}. In this paper, we use the DES Data Release 2 \citep[DR2][]{2018ApJS..239...18A,2021ApJS..255...20A}. \\

They provide shape measurements in their DR2 MAIN table, which is accessible through the Astro Data Lab (\url{https://datalab.noirlab.edu/des/index.php}).
They obtain these parameters applying \texttt{SourceExtractor} to the coadded images, which are combined from different epochs and different bands % ($r+i+z$) 
\citep[$r+i+z$,][]{2018PASP..130g4501M,2018ApJS..239...18A}.
We specifically use the \texttt{THETA\_J2000}, the position angle of the source in J2000 coordinates computed by \texttt{SourceExtractor} as explained above; and \texttt{ERRTHETA\_IMAGE}, its error. Their PA errors are higher than those of DESI LS, even though the camera they use is the same. This may be because DES relies on \texttt{SourceExtractor}, which fits a single ellipse, while DESI LS uses its own \texttt{Tractor} catalogue fitting more complex models. We also utilize the \texttt{EXTENDED\_CLASS\_COADD} parameter, based on the \texttt{SPREAD\-\_MODEL} parameter from \texttt{SourceExtractor}, which compares the PSF with a slightly extended circular exponential convolved with the PSF. Stars and point-like objects will have values of \texttt{SPREAD\_MODEL} very close to zero, while extended sources have higher values. \texttt{EXTENDED\_CLASS\_COADD} is defined as a sum of three Boolean conditions involving \texttt{SPREAD\_MODEL} in the i band. Every condition that is true adds one to the final value, so that

 \texttt{EXTENDED\_CLASS\_COADD} $= 0$ defines high-confidence stars, $\texttt{EXTENDED\_CLASS\_COADD} = 1$ defines likely stars, $\texttt{EXTENDED\_CLASS\_COADD} = 2$ defines likely galaxies and \texttt{EXTENDED\-\_CLASS\-\_COADD} $= 3$ defines high-confidence galaxies \citep{2021ApJS..254...24S}.

\subsection*{Underlying distribution of the galaxy and jet PAs}\label{sec:bias}
Before we start using the VLBI and optical PAs, we check for systematics in their distribution. 
While some surveys show a uniform distribution of PAs, others show non-uniform distributions, which may be caused by unaccounted for systematics.
We show the distribution of the VLBI jet PAs in Supplementary Figure \ref{fig:jetpa}, and as an example we show the two most extreme cases of the optical PAs in Supplementary Figure \ref{fig:opticalpa}: DESI LS with a very uniform distribution in panel (a), and SkyMapper with a completely skewed distribution in panel (b).

There are slightly more VLBI jets pointing towards the vertical direction (with a PA of either $0^{\circ}$ or $180^{\circ}$). \citet{2021A&A...653A.123M} show that this effect may be due to a North-South elongation effect from VLBI observations with an elliptical beam, especially affecting sources that are located in the southern region of the sky ($\text{DEC} < -30^{\circ}$). However, the PA measurements used in this paper were measured directly from the interferometric visibilities and therefore the PA measurements should not be affected in this way \citep{Plavin22}. The optical counterparts of SkyMapper present this strange preference for the same direction as well (an optical PA of $0^{\circ}$), while this effect was not seen in a random sample of galaxies from the same survey. \\

We describe how to account for this effect when quantifying the alignment signal below.

\subsection*{Errors}\label{sec:errors}
Because VLBI source positions are very precise (mas precision or better), we use their positions as the basis for cross-matching. We cross-match optical positions within a tolerance of one arcsecond (which is approximately equivalent to the maximum seeing limit of the optical surveys). This yields a total of 6273 sources that have optical cross-matches in at least one survey. \\

We obtain the errors in $\Delta$ PA by adding the error in the VLBI jet and in the optical axis PA in quadrature, assuming they are uncorrelated: \\
\begin{equation}
    \sigma_{\Delta{\text{PA}}} = \sqrt{\sigma_{\text{VLBI PA}}^2+\sigma_{\text{Optical PA}}^2}
\end{equation}
\\

To obtain the error bars of the bins in the $\Delta \text{PA}$ histograms, by assuming that the errors of the jet and galaxy PAs are Gaussian, we simulate their distributions by redrawing the PA in each survey 10000 times. 
We now call these errors $\sigma_{\Delta \text{PA}}^\text{MC}$ for Monte Carlo.  \\

\subsection*{Cuts}

\subsubsection*{Semi-minor axis cut}
We imposed a semi-minor axis $b$ cut keeping only sources with $b > 1.3"$ (`criterion 0) based on the mean DESI LS seeing value (See Table \ref{tab:listsurveys}) and applied it to all surveys. In general, this cut removes many sources from the `all cases' column, but also removes some sources in the `quality cuts', particularly in the `only with $z$' and $z<0.5$ rows, however, it does not affect the final results.

\subsubsection*{`Good case' cuts}
In the main text the `quality' and redshift cuts were described. 
In this section we provide some more relevant details.   \\

To define the `good case' cuts, we use various criteria from each optical survey to determine if an optical PA belongs to it', as listed in Table \ref{tab:listsurveys}
and defined as follows: \\

\begin{itemize}
    \item {\bf SDSS}: \texttt{class} = 3 sources (where 3 are extended and 6 are point-like) 
    \item {\bf DESI LS}: \texttt{type} = \texttt{EXP}, \texttt{DEV} and \texttt{SER} sources with observations in g,r, and z band and non-zero flux in all of the bands (where \texttt{EXP} is an exponential, \texttt{DEV} is a de Vaucouleurs and \texttt{SER} is a Sérsic model, while \texttt{PSF} is a point-like object).   
    \item {\bf SkyMapper}: minor axis $\texttt{b} > 2$ arcsec and stellar index $\texttt{CLASS\_STAR}<0.5$ (where $\texttt{CLASS\_STAR} = 0$ corresponds to ``not stars'' and $\texttt{CLASS\_STAR} = 1$ corresponds to ``stars'').
    \item {\bf KiDS}: minor axis $\texttt{B\_IMAGE}>2$ arcsec and stellar index $\texttt{CLASS\_STAR}<0.5$ (similarly to SkyMapper).
    \item {\bf DES}: \texttt{EXTENDED\_CLASS\_COADD} $\in \{2,3\}$ (where 2 corresponds to ``mostly galaxies'' and 3 to ``high confidence galaxies'').
\end{itemize}

In the case of the SDSS data, that all errors are set to $15^{\circ}$ means that there may be many unreliable PA measurements that passed the PA error cut. This is an additional motivation to use the `good case' cut. Sources that are cross-matched with optical counterparts and lack a spectroscopic redshift measurement are uploaded in an online repository. \\

\subsubsection*{Magnitude cuts}
We performed a magnitude cut for our DESI LS analysis. We performed cuts for $z < 23,20,17,14$ and we included the most severe ($z < 14$) in Supplementary Figure \ref{fig:DESI_mag_cuts}. The number of sources in the first row decreases significantly.  The second and third rows are identical in all cases. This tells us that all sources with redshifts and below magnitude $z<14$ have well measured shapes.  The final row is only marginally different. However, overall  the significances of the alignment signal remain consistent with Fig. \ref{fig:hist_desi}

\subsection*{PA comparisons}
It would seem that to visualize the alignment between VLBI jets and galaxies minor axis the best way is to check their correlation in a scatter plot. However, this method poses several problems.

\begin{figure*}[h!]             
    \includegraphics[width=\textwidth]{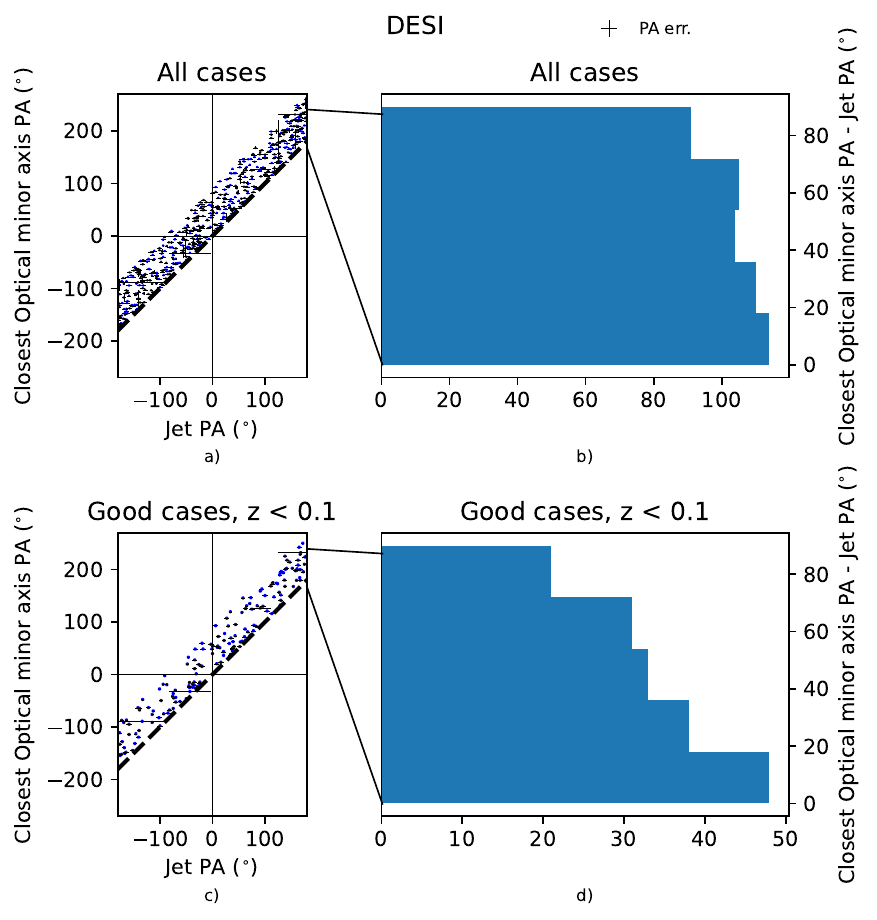}
    \caption{Optical PA vs. jet PA scatter plots and histograms for VLBI sources with a cross-match in DESI LS and a semi-minor axis greater than 1.3 arcsec. a) Scatter plot of the jet PA vs. the closest optical minor axis PA of all sources. The PA of the closest minor axis is taken so that it is always within a [0$^{\circ}$,90$^{\circ}$] range from the jet PA. The black dashed line shows a 1-1 relation. The closer a source is to the dashed line, the more aligned their jet and minor axis are. b) Histogram that shows the distribution of the difference between the optical PA and the jet PA (the vertical distance from any point to the diagonal). The same histogram as shown in the top left panel of Fig. \ref{fig:hist_desi} is recovered. c) Scatter plot of the jet PA vs. the closest optical minor axis PA of sources within the `good case' cut and a redshift below 0.1. The PA of the closest minor axis is taken so that it is always within a [0$^{\circ}$,90$^{\circ}$] range from the jet PA. The black dashed line shows a 1-1 relation. The closer a source is to the dashed line, the more aligned their jet and minor axis are. d) Histogram that shows the distribution of the difference between the optical PA and the jet PA (the vertical distance from any point to the diagonal). The same histogram as shown in the bottom right panel of Fig. \ref{fig:hist_desi} is recovered.}
    \label{fig:DESI_scatterplots}
\end{figure*}

The main issue is that this plot would not take the circular nature of the angles into account. The visualization of the angular distance between two directions in a linear plot can get tricky if one imposes the wrong constraints to the range of each angle. We especially have to consider that the semi-minor axis has a 180$^{\circ}$ rotational symmetry, while the jet has one specific direction. The VLBI jet PA is defined within a [-180$^{\circ}$,180$^{\circ}$] range, to cover to full range of directions, while the galactic minor axis is within a [-90$^{\circ}$,90$^{\circ}$] range, which is enough due to its symmetry. If we made a scatter plot with these default ranges, there would be a misleading representation of the true alignment. If, for example, a jet has a PA = 87$^{\circ}$ and the minor axis has a PA =  -87$^{\circ}$, the point would fall far away from the diagonal (in one of the corners in the scatter plot), while the real angular difference between the two is $\Delta$PA = 6$^{\circ}$, very close to aligned. To handle this, we take the PA of the semi-minor axis and add or subtract 180$^{\circ}$ when needed until its numerical value is always within 
a [0$^{\circ}$,90$^{\circ}$] separation from the numerical value of the jet PA. In the previous example, the semi-minor axis PA would become 93$^{\circ}$. This turns the semi-minor axis range to [-270$^{\circ}$,270$^{\circ}$], and the points will always fall within a +90$^{\circ}$ band of the diagonal.
Supplementary Figure \ref{fig:DESI_scatterplots} shows this computation for two examples: one with all sources with semi-minor axis $b > 1.3"$ where there is no signal, and one with the good case with $b > 1.3"$ and $z < 0.5$. The important point is that we can directly recover the $\Delta$PA histograms (on the right) from the scatter plots (on the left) by taking $\Delta\text{PA} = y-x$. We have illustrated that by putting the histograms at the right of the scatter plots.

However, even here, we see that the alignment signal is not strong enough for the correlation to be easily visible. To aid with the visualization, we make a coloured 2D histogram in Supplementary Figure \ref{fig:DESI_scatter_hist} to help show the excess (or lack of excess) of sources in a given bin.

\subsection*{Quantifying the alignment signal}
\label{sec:systematics}

The underlying distributions of PAs of some of the surveys may not be uniform.
This lack of uniformity might influence the results and even create a false alignment signal. We want to make sure that the signal we find is not affected by this. In order to do that, we study how far the observed signal is from what we would see if the PAs were taken at random from their underlying distributions. 
This approach is commonly used to detect alignment signals\cite{brainerd2005,lhuillier2017}. \\

\begin{figure*}[h!]
    \includegraphics[width=\textwidth]{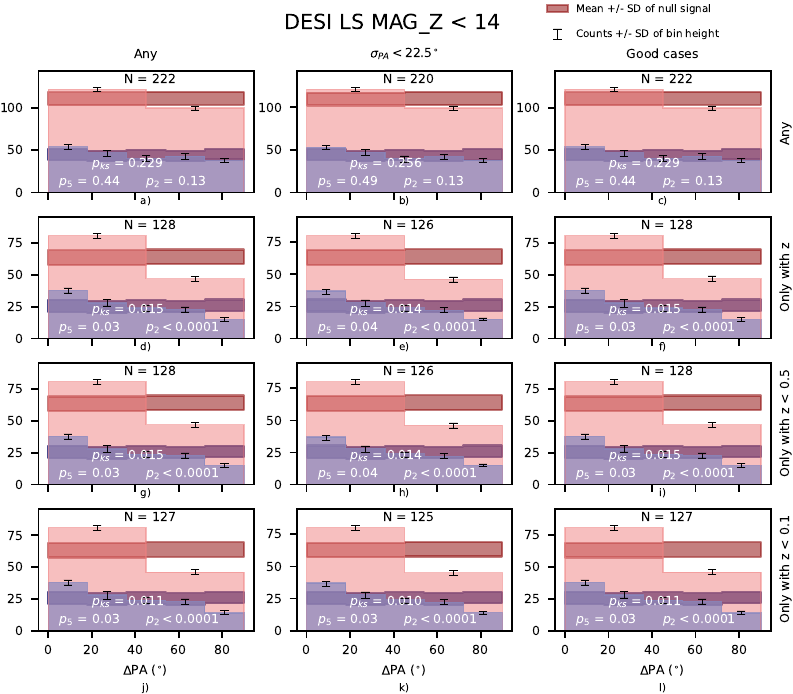}
    \caption{All VLBI sources with a cross-match in DESI LS, a semi-minor axis $>$ 1.3" and a magnitude in the $z$ band $<14$ (the `criterion 1'). $\Delta \text{PA}$ indicates the angle between the VLBI jet and the optical minor axis of its host galaxy. The dark bands show the expected null signal (no alignment), which is the result of our careful analysis of the systematic errors. The stronger the deviation from the dark band the stronger the alignment signal. $p_{10}$ and $p_2$ are the p-values for ten bins and two bins in the $\Delta\text{PA}$ space respectively. $p_{\text{ks}}$ is the p-value from a KS test comparing the data against a uniform distribution, independently of the number of bins. Each panel corresponds to a specific selection of sources, showing the number of sources on top. Each panel corresponds to a specific selection of sources based on several cuts. The number of sources inside each panel is shown on top. a) All `criterion 1' sources. b) All `criterion 1' sources whose jet and optical PA error is $<$ $22.5^{\circ}$. c) All `criterion 1' sources within the `good case'. d) All `criterion 1' sources with a measured spectroscopic redshift. e) All `criterion 1' sources whose jet and optical PA error is $<$ $22.5^{\circ}$ and with a measured spectroscopic redshift. f) All `criterion 1' sources within the `good case' and with a measured spectroscopic redshift. g) All `criterion 1' sources with $z < 0.5$. h) All `criterion 1' sources whose jet and optical PA error is $<$ $22.5^{\circ}$ and with $z < 0.5$. i) All `criterion 1' sources within the `good case' and with $z < 0.5$. j) All `criterion 1' sources with $z < 0.1$ k) All `criterion 1' sources whose jet and optical PA error is $<$ $22.5^{\circ}$ and with $z < 0.1$. l) All `criterion 1' sources within the `good case' and with $z < 0.1$.}
    \label{fig:DESI_mag_cuts}
\end{figure*}

This analysis is applicable to both the two and  five bin case. In order to do this, we shuffle the underlying PA distributions of both the jets and the optical minor axes $N_\text{RS} = 1000$ times. For each reshuffling (RS) $j$, we recalculate the histogram of the reshuffled PA angle differences, $\Delta\text{PA}^\text{RS,j}$.  The height of each histogram bin after doing the reshuffling is $N_{i,j}$, where $i$ denotes the bin and $j$ the reshuffling. For each bin $i$ we  calculate the mean ($\bar N_{\text{RS},i}$) and the standard deviation ($\sigma_{\text{RS},i}$) of the number counts from all the reshufflings, defined as  
\begin{align}
   \bar N_{\text{RS},i}& =\frac 1 {N_\text{RS}} \sum_{j=1}^{N_\text{RS}} N_{i,j}  \\
\sigma^2_{\text{RS},i}& =\frac 1 {N_\text{RS}} \sum_{j=1}^{N_\text{RS}} (N_{i,j} - \bar N_{\text{RS},i})^2,
\end{align}
where $\bar N_{\text{RS},i}$ is the center of the dark bands and $\sigma_{\text{RS},i}$ is the (vertical) width of those bands present in each bin of the histograms in Fig. \ref{fig:hist_desi} and Supplementary Figures \ref{fig:combined_histogram}-\ref{fig:DESI_mag_cuts}. These bands, as explained, reflect the distribution of the underlying data. 
Therefore, if there are systematics in the data this would be reflected as a deviation from a flat band.
 \\

Next, we define a  statistic ($Q$) to quantify the significance of any possible signal away from a random occurrence.  On one hand, we compute  $Q$ using the bin heights of the original, unshuffled histograms ($N_{\mathrm{data,i}}$) and comparing them with the mean and standard deviation of the reshuffled data (the dark bands) as defined before. This is called $Q_{\mathrm{data}}$: \\

\begin{equation}
Q_{\mathrm{data}}=\frac{1} {N_\mathrm{bins}} 
\sum_{i=1}^{N_\mathrm{bins}}
\left[
\frac 
{N_{\mathrm{data},i}-\bar N_{\mathrm{RS},i}} 
{\sigma_{\mathrm{RS},i}}
  \right]^2,
\end{equation}
%where $N_{\mathrm{data},i}$ is the measured number of jet-galaxy pairs in the bin. 
$Q_\mathrm{data}$ is thus a quantification of the deviation from the expected random signal as computed by reshuffling.
A low $Q_\mathrm{data}$ corresponds to consistency between the measured and the randomized signals, while a true alignment signal would yield a large $Q_\mathrm{data}$.

In order to determine the statistical significance we calculate  $Q$ for each of the 1000 reshufflings $j$ as:
\begin{equation}
    Q_j = \frac{1} {N_\mathrm{bins}} 
\sum_{i=1}^{N_\mathrm{bins}}
\left[
\frac 
{N_{j,i}-\bar N_{\mathrm{RS},i}} 
{\sigma_{\mathrm{RS},i}}
  \right]^2.
\end{equation}
where $N_{j,i}$ is the height of bin $i$ in each reshuffling $j$.   

We now calculate the $p$-value, i.e, the probability  that any given $Q_j$ is greater than $Q_\text{data}$ under the null hypothesis $H_0$ of non-alignment $P(Q_j>Q_\text{data}|H_0)$. This shows how far the $Q_\text{data}$ lies from the distribution of the 1000 $Q_j$ from all the reshufflings.

\subsection*{Effect of binning}
We investigate the behavior of the results using different numbers of bins (2,4,5,6,8,10) for the histograms. We found that even for higher bin numbers, the p-values remain consistent, sometimes increasing and sometimes decreasing but not changing the overall picture. We present the most severe case of 10 bins (and the 2-bin case as a reference) in Supplementary Figure \ref{fig:DESI_10_bins}.

\subsection*{Redshift evolution}
\begin{figure*}[h!]
    \includegraphics[width=\textwidth]{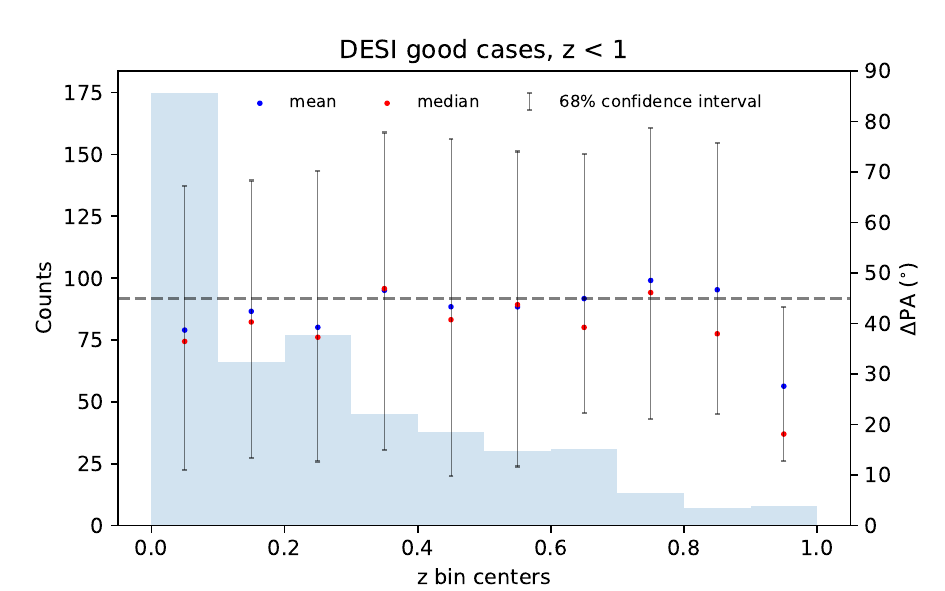}
    \caption{Median and mean $\Delta$PA for each redshift bin for DESI LS good case sources with z $<$ 1. The error bars show the 68\% confidence interval. The dashed line shows $\Delta \text{PA} = 45^\circ$. The histogram in the background shows the number of sources in each bin.}
    \label{fig:DESI_delta_redshift}
\end{figure*}

We show the median $\Delta$PA vs. redshift of the DESI `good cases' with a redshift z $<$ 1 in Fig. \ref{fig:DESI_delta_redshift}, generating redshift bins and taking the median $\Delta$PA of the sources in each bin. We overplot the number of sources inside each bin. We find that in general that both the mean and median $\Delta$PA are below 45$^{\circ}$, particularly in redshift bins with high source counts. We find no clear trend with redshift. Overall the median and 68\% errors are ${40^{+34}_{-27}}^{\circ}$

\subsection*{Eagle simulation projection effects}

To investigate how projection effects could affect the observed 2D $\Delta$PA distribution, we use data from the 100 Mpc box of EAGLE simulation \citep{EAGLE_sim_1,EAGLE_sim_2}. This cosmological hydrodynamical simulation provides catalogs \citep{EAGLE_catalogs} for a wide range of properties of their galaxies. This includes the morphology and kinematics of their galaxies \citep{Thob:2019}, profile fits to artificial images \citep{deGraaff:2022} created from their simulated galaxies that provide position the position angles along the z-axis of the simulation as well as colours and luminosity \citep{Trayford:2015} of said galaxies. Furthermore, the EAGLE simulation provides the 3D spin vectors of several components of the galaxies (stars, cold and hot gas).

The 3D stellar spin vector is assumed to be close to perpendicular to the main plane of the galaxy. To confirm this, we split the sample into ellipticals and spirals using a cut in the u-r color ($u-r > 2.2$ for ellipticals, $u-r < 2.2$ for spirals). We see no significant differences in the results from both samples, but nevertheless use the elliptical sample. To emphasize, we are not implying a physical connection between the stellar spin and the jet direction, we are just investigating projection effects. We then project the stellar spin onto the same plane as the projected $PA$ of the galaxies and compute the 2D projected $\Delta \text{PA}$ between the minor axis and the spin, producing histograms in a similar style to Fig. \ref{fig:hist_desi}.

We study 4 different scenarios, as mentioned in the main text.
i) Simulation data with no additional errors.
ii) Fully randomized stellar spin directions.
iii) and iv) Added scatter to the spin directions. We provide more details below.

To account for differences between a `pure' and a more realistic connection, we introduce a scatter in the 3D spins before projecting them. We draw angles from both a Gaussian (iii) and a uniform (iv) distribution for $\theta$, where $\theta \in N(0^{\circ}, \epsilon \cdot 90^{\circ})$ and $\theta \in U(0,\epsilon \cdot 90^{\circ})$, and a uniform distribution for $\phi$ in both cases, where $\phi \in U(0, 360)$. Here, $\epsilon \in [0, 1]$ represents the fractional scatter. We then tilt the spin by an amplitude angle $\theta$ and a random orientation $\phi$. Our dataset consists of approximately 3000 galaxies from the simulation. To generate a $\Delta$PA histogram, we repeat the process 1000 times, each time sampling the angles randomly from the distributions. Finally, we calculate the average and standard deviation for each bin in the $\Delta$PA space and plot the final histogram. We show the resulting histograms in Fig. \ref{fig:Eagle_sim}.

\subsection*{Combined histogram}
If the VLBI source is cross-matched to multiple optical surveys, we compute the weighted average of the angle difference ($\overline{\Delta{\text{PA}}}$) and its error ($\sigma_{\overline{\Delta{\text{PA}}}}$) in quadrature:

\begin{align}
    \overline{\Delta{\text{PA}}} & = \sum_i{w_i \, {\Delta{\text{PA}}}_i},  \\
    \sigma_{\overline{\Delta{\text{PA}}}} & = \sqrt{\frac{1}{\sum_i{{\sigma^{-2}_{\Delta{\text{PA}}}}}_i}},
\end{align}
where $i$ refers to the optical survey and $w_i$, ${\Delta{\text{PA}}}_i$ and $\sigma_{\Delta{\text{PA}}_i}$ are the corresponding weight, angle difference and its error. 
The weights are defined as:
\begin{equation}
    w_i = \frac{1/\sigma_{\Delta{\text{PA}}_i}^2}{\sum_i{1/\sigma_{\Delta{\text{PA}}_i}^2}}.
\end{equation}

In Supplementary Figure \ref{fig:combined_histogram} we present the distributions of the weighted angle between the VLBI jet PA and the minor axis of the apparent orientation of the host galaxy ($\overline{\Delta{\text{PA}}}$). The histogram is divided in the same way as described in the main text (e.g. Fig. \ref{fig:hist_desi}). \\

In general, the combined histogram follows the DESI LS histogram in Fig. \ref{fig:hist_desi}, as described in the main text.  As before, we can see several trends in the data. In the left column containing all sources we don't see any alignment overall ($0.3 < p < 0.5$), but similarly to the DESI LS histogram (Fig. \ref{fig:hist_desi}) we see a weak alignment for $z<0.1$. When we look at the `good' cases, we begin to see a trend towards $\overline{\Delta{\text{PA}}} = 0^{\circ}$ (i.e. the inner jet PA aligned with the minor axis of the projected optical orientation). This trend becomes slightly more significant if the sources have spectroscopic redshifts ($0.1 < p < 0.3$), and is especially strong for sources with spectroscopic $z<0.1$ ($0.01 < p < 0.1$) 

\subsection* {Survey by survey analysis}\label{sec:survey}

We also perform the analysis on a survey-by-survey basis and show the results in Supplementary Figures \ref{fig:hist_sdss}-\ref{fig:hist_SM}. We do not include the individual analysis of the KiDS catalogue here due to the very low number of cross-matches. \\

\section*{Data availability}
The data and codes for reproducing the results of this work are available via GitHub at \url{https://github.com/davfer12/Detection-of-an-orthogonal-alignment-between-parsec-scale-AGN-jets-and-their-host-galaxies}. In the repository, we provide data for the cross-matched sources, but we do not include the whole catalogues as they are publicly available from their own repositories.

\section* {Acknowledgements}
Authors want to acknowledge useful discussions with Dimitry Blinov, % Françoise Combes, 
Iannis Liodakis, I. Sevilla-Noarbe, and Dustin Lang.  
Authors also want to acknowledge the Friends of Sejong 2023 conference \url{https://sites.google.com/view/friends-of-sejong2024} 
in which most of the results were discussed.  

D. F. G. and J. A. H. acknowledges the support of the National Research Foundation of Korea (NRF-2021R1C1C1009973).

B.~L. acknowledges the support of the National Research Foundation of Korea (NRF-2022R1F1A1076338) and the support of the Korea Institute for Advanced Study (KIAS) grant funded by the government of Korea.

D. F. G. acknowledges financial support by the Predoctoral Gobierno de Aragón fellowship 2023-2027.

J. A. acknowledges the support of the Spanish Ministerio de Universidades and European Union NextGenerationEU through the Mar\'ia Zambrano program (CT33/21) at Universidad Complutense de Madrid (UCM), UCM project PR3/23-30808 and the MICINN (Spain) project PID2022-138263NB-I0 (AEI/FEDER, UE).

C.S. acknowledges support from the National Research Foundation of Korea (NRF) through grant No. 2021R1A2C101302413 funded by the Korean Ministry of Education, Science and Technology (MoEST). 

M.J.J. acknowledges the support from the NRF through grant Nos. 2022R1A2C1003130 and RS-2023-00219959.

D.P. acknowledges support from the project ``Understanding Dark Universe Using Large Scale Structure of the Universe'', funded by the Korean Ministry of Science.
\\
We acknowledge the use of the Astrogeo VLBI FITS image database at \url{http://astrogeo.smce.nasa.gov/vlbi_images} maintained by Leonid Petrov.
\\
The Legacy Surveys consist of three individual and complementary projects: the Dark Energy Camera Legacy Survey (DECaLS; Proposal ID \#2014B-0404; PIs: David Schlegel and Arjun Dey), the Beijing-Arizona Sky Survey (BASS; NOAO Prop. ID \#2015A-0801; PIs: Zhou Xu and Xiaohui Fan), and the Mayall z-band Legacy Survey (MzLS; Prop. ID \#2016A-0453; PI: Arjun Dey). DECaLS, BASS and MzLS together include data obtained, respectively, at the Blanco telescope, Cerro Tololo Inter-American Observatory, NSF’s NOIRLab; the Bok telescope, Steward Observatory, University of Arizona; and the Mayall telescope, Kitt Peak National Observatory, NOIRLab. Pipeline processing and analyses of the data were supported by NOIRLab and the Lawrence Berkeley National Laboratory (LBNL). The Legacy Surveys project is honored to be permitted to conduct astronomical research on Iolkam Du’ag (Kitt Peak), a mountain with particular significance to the Tohono O’odham Nation.
\\
NOIRLab is operated by the Association of Universities for Research in Astronomy (AURA) under a cooperative agreement with the National Science Foundation. LBNL is managed by the Regents of the University of California under contract to the U.S. Department of Energy.
\\
This project used data obtained with the Dark Energy Camera (DECam), which was constructed by the Dark Energy Survey (DES) collaboration. Funding for the DES Projects has been provided by the U.S. Department of Energy, the U.S. National Science Foundation, the Ministry of Science and Education of Spain, the Science and Technology Facilities Council of the United Kingdom, the Higher Education Funding Council for England, the National Center for Supercomputing Applications at the University of Illinois at Urbana-Champaign, the Kavli Institute of Cosmological Physics at the University of Chicago, Center for Cosmology and Astro-Particle Physics at the Ohio State University, the Mitchell Institute for Fundamental Physics and Astronomy at Texas A\&M University, Financiadora de Estudos e Projetos, Fundacao Carlos Chagas Filho de Amparo, Financiadora de Estudos e Projetos, Fundacao Carlos Chagas Filho de Amparo a Pesquisa do Estado do Rio de Janeiro, Conselho Nacional de Desenvolvimento Cientifico e Tecnologico and the Ministerio da Ciencia, Tecnologia e Inovacao, the Deutsche Forschungsgemeinschaft and the Collaborating Institutions in the Dark Energy Survey. The Collaborating Institutions are Argonne National Laboratory, the University of California at Santa Cruz, the University of Cambridge, Centro de Investigaciones Energeticas, Medioambientales y Tecnologicas-Madrid, the University of Chicago, University College London, the DES-Brazil Consortium, the University of Edinburgh, the Eidgenossische Technische Hochschule (ETH) Zurich, Fermi National Accelerator Laboratory, the University of Illinois at Urbana-Champaign, the Institut de Ciencies de l'Espai (IEEC/CSIC), the Institut de Fisica d'Altes Energies, Lawrence Berkeley National Laboratory, the Ludwig Maximilians Universitat Munchen and the associated Excellence Cluster Universe, the University of Michigan, NSF’s NOIRLab, the University of Nottingham, the Ohio State University, the University of Pennsylvania, the University of Portsmouth, SLAC National Accelerator Laboratory, Stanford University, the University of Sussex, and Texas A\&M University.
\\
BASS is a key project of the Telescope Access Program (TAP), which has been funded by the National Astronomical Observatories of China, the Chinese Academy of Sciences (the Strategic Priority Research Program “The Emergence of Cosmological Structures” Grant \# XDB09000000), and the Special Fund for Astronomy from the Ministry of Finance. The BASS is also supported by the External Cooperation Program of Chinese Academy of Sciences (Grant \# 114A11KYSB20160057), and Chinese National Natural Science Foundation (Grant \# 12120101003, \# 11433005).
\\
The Legacy Survey team makes use of data products from the Near-Earth Object Wide-field Infrared Survey Explorer (NEOWISE), which is a project of the Jet Propulsion Laboratory/California Institute of Technology. NEOWISE is funded by the National Aeronautics and Space Administration.
\\
The Legacy Surveys imaging of the DESI LS footprint is supported by the Director, Office of Science, Office of High Energy Physics of the U.S. Department of Energy under Contract No. DE-AC02-05CH1123, by the National Energy Research Scientific Computing Center, a DOE Office of Science User Facility under the same contract; and by the U.S. National Science Foundation, Division of Astronomical Sciences under Contract No. AST-0950945 to NOAO.
\\
Funding for the Sloan Digital Sky Survey IV has been provided by the Alfred P. Sloan Foundation, the U.S. Department of Energy Office of Science, and the Participating Institutions. 
\\
SDSS-IV acknowledges support and resources from the Center for High Performance Computing  at the University of Utah. The SDSS website is \url{www.sdss4.org}.
\\
SDSS-IV is managed by the Astrophysical Research Consortium for the Participating Institutions of the SDSS Collaboration including the Brazilian Participation Group, the Carnegie Institution for Science, Carnegie Mellon University, Center for Astrophysics | Harvard \& Smithsonian, the Chilean Participation Group, the French Participation Group, Instituto de Astrof\'isica de Canarias, The Johns Hopkins University, Kavli Institute for the Physics and Mathematics of the Universe (IPMU) / University of Tokyo, the Korean Participation Group, Lawrence Berkeley National Laboratory, Leibniz Institut f\"ur Astrophysik Potsdam (AIP),  Max-Planck-Institut f\"ur Astronomie (MPIA Heidelberg), Max-Planck-Institut f\"ur Astrophysik (MPA Garching), Max-Planck-Institut f\"ur Extraterrestrische Physik (MPE), National Astronomical Observatories of China, New Mexico State University, New York University, University of Notre Dame, Observat\'ario Nacional / MCTI, The Ohio State University, Pennsylvania State University, Shanghai Astronomical Observatory, United Kingdom Participation Group, Universidad Nacional Aut\'onoma de M\'exico, University of Arizona, University of Colorado Boulder, University of Oxford, University of Portsmouth, University of Utah, University of Virginia, University of Washington, University of Wisconsin, Vanderbilt University, and Yale University.
\\
This project used public archival data from the Dark Energy Survey (DES). Funding for the DES Projects has been provided by the U.S. Department of Energy, the U.S. National Science Foundation, the Ministry of Science and Education of Spain, the Science and Technology FacilitiesCouncil of the United Kingdom, the Higher Education Funding Council for England, the National Center for Supercomputing Applications at the University of Illinois at Urbana-Champaign, the Kavli Institute of Cosmological Physics at the University of Chicago, the Center for Cosmology and Astro-Particle Physics at the Ohio State University, the Mitchell Institute for Fundamental Physics and Astronomy at Texas A\&M University, Financiadora de Estudos e Projetos, Funda{\c c}{\~a}o Carlos Chagas Filho de Amparo {\`a} Pesquisa do Estado do Rio de Janeiro, Conselho Nacional de Desenvolvimento Cient{\'i}fico e Tecnol{\'o}gico and the Minist{\'e}rio da Ci{\^e}ncia, Tecnologia e Inova{\c c}{\~a}o, the Deutsche Forschungsgemeinschaft, and the Collaborating Institutions in the Dark Energy Survey.
\\
The Collaborating Institutions are Argonne National Laboratory, the University of California at Santa Cruz, the University of Cambridge, Centro de Investigaciones Energ{\'e}ticas, Medioambientales y Tecnol{\'o}gicas-Madrid, the University of Chicago, University College London, the DES-Brazil Consortium, the University of Edinburgh, the Eidgen{\"o}ssische Technische Hochschule (ETH) Z{\"u}rich,  Fermi National Accelerator Laboratory, the University of Illinois at Urbana-Champaign, the Institut de Ci{\`e}ncies de l'Espai (IEEC/CSIC), the Institut de F{\'i}sica d'Altes Energies, Lawrence Berkeley National Laboratory, the Ludwig-Maximilians Universit{\"a}t M{\"u}nchen and the associated Excellence Cluster Universe, the University of Michigan, the National Optical Astronomy Observatory, the University of Nottingham, The Ohio State University, the OzDES Membership Consortium, the University of Pennsylvania, the University of Portsmouth, SLAC National Accelerator Laboratory, Stanford University, the University of Sussex, and Texas A\&M University.
\\
Based in part on observations at Cerro Tololo Inter-American Observatory, National Optical Astronomy Observatory, which is operated by the Association of Universities for Research in Astronomy (AURA) under a cooperative agreement with the National Science Foundation.
\\
The national facility capability for SkyMapper has been funded through ARC LIEF grant LE130100104 from the Australian Research Council, awarded to the University of Sydney, the Australian National University, Swinburne University of Technology, the University of Queensland, the University of Western Australia, the University of Melbourne, Curtin University of Technology, Monash University and the Australian Astronomical Observatory. SkyMapper is owned and operated by The Australian National University's Research School of Astronomy and Astrophysics. The survey data were processed and provided by the SkyMapper Team at ANU. The SkyMapper node of the All-Sky Virtual Observatory (ASVO) is hosted at the National Computational Infrastructure (NCI). Development and support of the SkyMapper node of the ASVO has been funded in part by Astronomy Australia Limited (AAL) and the Australian Government through the Commonwealth's Education Investment Fund (EIF) and National Collaborative Research Infrastructure Strategy (NCRIS), particularly the National eResearch Collaboration Tools and Resources (NeCTAR) and the Australian National Data Service Projects (ANDS)
\\
We acknowledge the Virgo Consortium for making their simulation data available. The EAGLE simulations were performed using the DiRAC-2 facility at Durham, managed by the ICC, and the PRACE facility Curie based in France at TGCC, CEA, Bruy\`{e}res-le-Ch\^{a}tel.

\subsection* {Software:} This research made use of Astropy, a community-developed core Python package for Astronomy \citep{astropy:2013, astropy:2018}, the HEALPix and Healpy package \citep{2005ApJ...622..759G,Zonca2019}, the  Numpy package \cite{book}, the Scipy package \citep{2019arXiv190710121V} and Matplotlib package \citep{Hunter:2007}.

\section*{Author Contributions Statement}
D. F. G. lead and developed the project, produced the code and figures, made the data analysis and wrote the paper. J. H. designed and lead the project and wrote the paper. B. L. helped with the design of the project, did the statistical analysis and wrote the paper. J. A. provided help with the optical catalogues and contributed to the data analysis and the paper writing. C. S. prepared and assembled the catalogues of imaging data, the simulation, and co-wrote the sections on them in the paper. K. F. provided guidance to the first author on the intricacies of galaxy shape measurements and the writing of the text. M. J. J. helped with the measurement of galaxy orientation.  D. P. helped with the design of the project. F. C. helped to interpret the results and write justification concerning the possible alignment or misalignment between the small scale radio-jets and the optical spin axis of the galaxies.

\section*{Competing Interests}
The authors declare no competing interests

\newpage

% this is the bbl as of 2024.09.27: 

\newpage

\section*{Supplementary Materials}

\begin{figure}[h]
    \centering
    \includegraphics[width=0.5\textwidth]{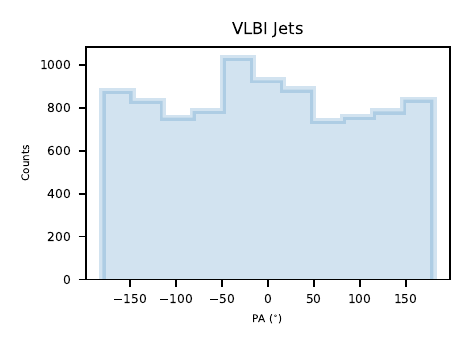}
    \caption{Histogram of VLBI jet PAs. The PAs come from the Astrogeo catalogue and vary between -180$^{\circ}$ and 180$^{\circ}$.}
    \label{fig:jetpa}
\end{figure}

\begin{figure}
    \centering
    \includegraphics[width=\textwidth]{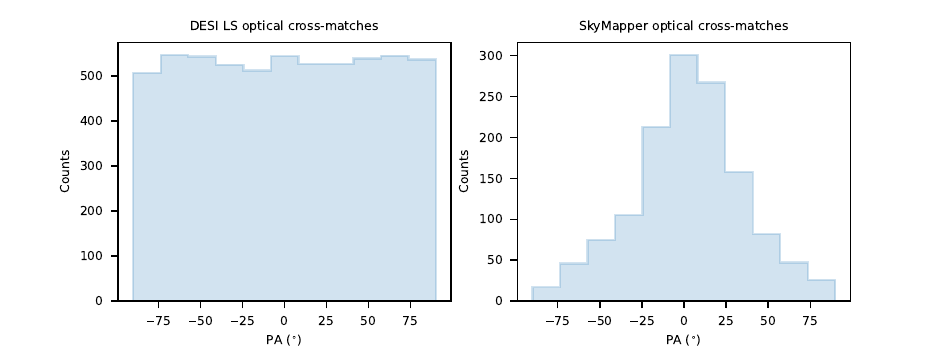}
    \caption{PA distribution of the cross-matches with two of the optical surveys. a) DESI LS shows the most uniform distribution. b) SkyMapper shows the least uniform distribution.}
    \label{fig:opticalpa}
\end{figure}

\begin{figure*}[h!]
    \includegraphics[width=\textwidth]{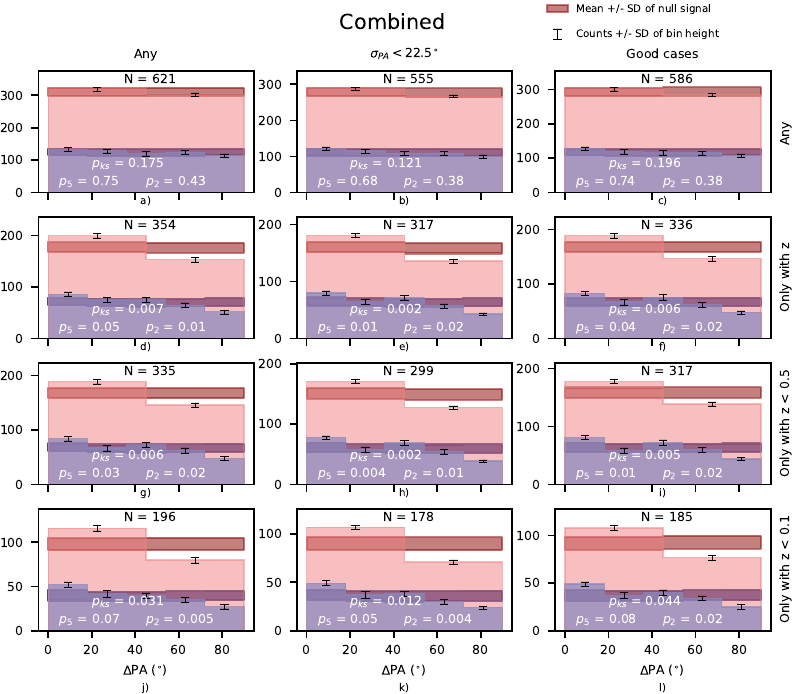}
    \caption{
     All VLBI sources with at least one optical cross-match and a semi-minor axis $>$ 1.3" (the `criterion 0'). If they are present in several surveys, the weighted average of the angle difference is taken, using the optical and VLBI errors of the PAs as weights. 
    $\Delta \text{PA}$ indicates the angle between the VLBI jet and the optical minor axis of its host galaxy. 
    The dark bands show the expected null signal (no alignment), which is the result of our careful analysis of the systematic errors. The stronger the deviation from the dark band the stronger the alignment signal. $p_5$ and $p_2$ are the p-values for five bins and two bins in the $\Delta\text{PA}$ space respectively. $p_{\text{ks}}$ is the p-value from a KS test comparing the data against a uniform distribution, independently of the number of bins. Each panel corresponds to a specific selection of sources, showing the number of sources on top. Each panel corresponds to a specific selection of sources based on several cuts. The number of sources inside each panel is shown on top. a) All `criterion 0' sources. b) All `criterion 0' sources whose jet and optical PA error is $<$ $22.5^{\circ}$. c) All `criterion 0' sources within the `good case'. d) All `criterion 0' sources with a measured spectroscopic redshift. e) All `criterion 0' sources whose jet and optical PA error is $<$ $22.5^{\circ}$ and with a measured spectroscopic redshift. f) All `criterion 0' sources within the `good case' and with a measured spectroscopic redshift. g) All `criterion 0' sources with $z < 0.5$. h) All `criterion 0' sources whose jet and optical PA error is $<$ $22.5^{\circ}$ and with $z < 0.5$. i) All `criterion 0' sources within the `good case' and with $z < 0.5$. j) All `criterion 0' sources with $z < 0.1$ k) All `criterion 0' sources whose jet and optical PA error is $<$ $22.5^{\circ}$ and with $z < 0.1$. l) All `criterion 0' sources within the `good case' and with $z < 0.1$.
    }    \label{fig:combined_histogram}
\end{figure*}

\begin{figure*}[h]
    \includegraphics[width=\textwidth]{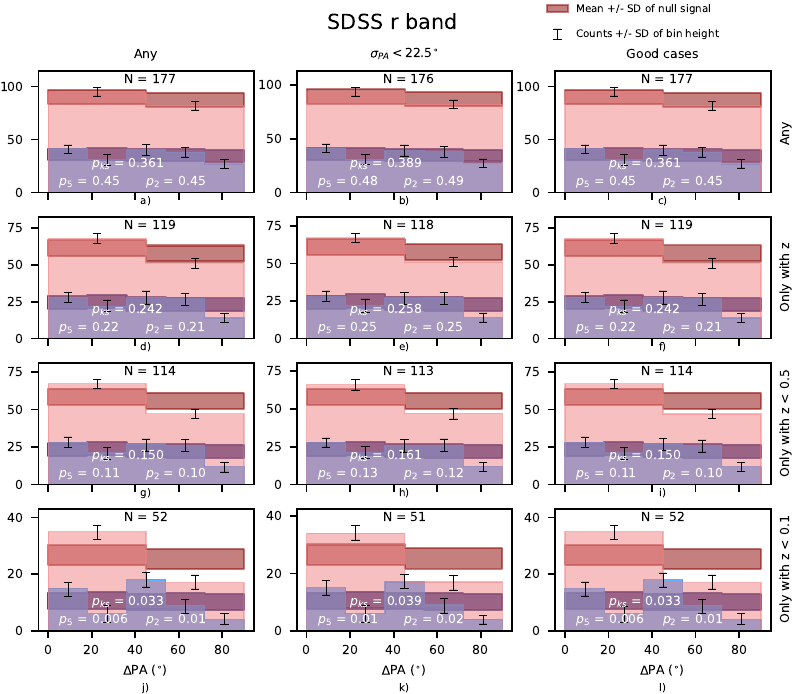}
    \caption{
    All VLBI sources with a cross-match in SDSS and a semi-minor axis $>$ 1.3" (the `criterion 0'). $\Delta \text{PA}$ indicates the angle between the VLBI jet and the optical minor axis of its host galaxy. The dark bands show the expected null signal (no alignment), which is the result of our careful analysis of the systematic errors. The stronger the deviation from the dark band the stronger the alignment signal. $p_5$ and $p_2$ are the p-values for five bins and two bins in the $\Delta\text{PA}$ space respectively. $p_{\text{ks}}$ is the p-value from a KS test comparing the data against a uniform distribution, independently of the number of bins. Each panel corresponds to a specific selection of sources, showing the number of sources on top. Each panel corresponds to a specific selection of sources based on several cuts. The number of sources inside each panel is shown on top. a) All `criterion 0' sources. b) All `criterion 0' sources whose jet and optical PA error is $<$ $22.5^{\circ}$. c) All `criterion 0' sources within the `good case'. d) All `criterion 0' sources with a measured spectroscopic redshift. e) All `criterion 0' sources whose jet and optical PA error is $<$ $22.5^{\circ}$ and with a measured spectroscopic redshift. f) All `criterion 0' sources within the `good case' and with a measured spectroscopic redshift. g) All `criterion 0' sources with $z < 0.5$. h) All `criterion 0' sources whose jet and optical PA error is $<$ $22.5^{\circ}$ and with $z < 0.5$. i) All `criterion 0' sources within the `good case' and with $z < 0.5$. j) All `criterion 0' sources with $z < 0.1$ k) All `criterion 0' sources whose jet and optical PA error is $<$ $22.5^{\circ}$ and with $z < 0.1$. l) All `criterion 0' sources within the `good case' and with $z < 0.1$.
    }
    \label{fig:hist_sdss}
\end{figure*}

\begin{figure*}[h]
    \includegraphics[width=\textwidth]{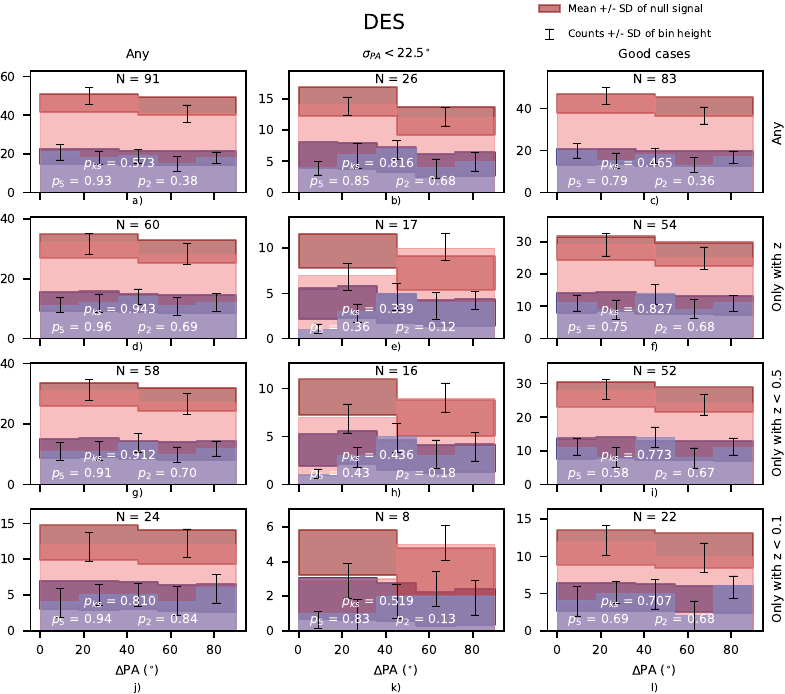}
    \caption{All VLBI sources with a cross-match in DES and a semi-minor axis $>$ 1.3" (the `criterion 0'). $\Delta \text{PA}$ indicates the angle between the VLBI jet and the optical minor axis of its host galaxy. The dark bands show the expected null signal (no alignment), which is the result of our careful analysis of the systematic errors. The stronger the deviation from the dark band the stronger the alignment signal. $p_5$ and $p_2$ are the p-values for five bins and two bins in the $\Delta\text{PA}$ space respectively. $p_{\text{ks}}$ is the p-value from a KS test comparing the data against a uniform distribution, independently of the number of bins. Each panel corresponds to a specific selection of sources, showing the number of sources on top. Each panel corresponds to a specific selection of sources based on several cuts. The number of sources inside each panel is shown on top. a) All `criterion 0' sources. b) All `criterion 0' sources whose jet and optical PA error is $<$ $22.5^{\circ}$. c) All `criterion 0' sources within the `good case'. d) All `criterion 0' sources with a measured spectroscopic redshift. e) All `criterion 0' sources whose jet and optical PA error is $<$ $22.5^{\circ}$ and with a measured spectroscopic redshift. f) All `criterion 0' sources within the `good case' and with a measured spectroscopic redshift. g) All `criterion 0' sources with $z < 0.5$. h) All `criterion 0' sources whose jet and optical PA error is $<$ $22.5^{\circ}$ and with $z < 0.5$. i) All `criterion 0' sources within the `good case' and with $z < 0.5$. j) All `criterion 0' sources with $z < 0.1$ k) All `criterion 0' sources whose jet and optical PA error is $<$ $22.5^{\circ}$ and with $z < 0.1$. l) All `criterion 0' sources within the `good case' and with $z < 0.1$.}
    \label{fig:hist_des}
\end{figure*}

\begin{figure*}[h]
    \includegraphics[width=\textwidth]{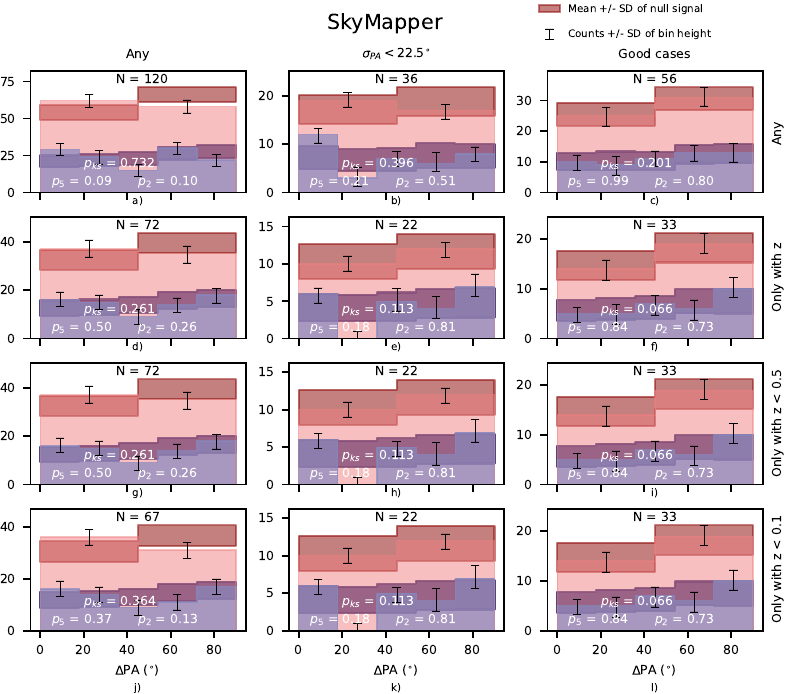}
    \caption{All VLBI sources with a cross-match in Skymapper and a semi-minor axis $>$ 1.3" (the `criterion 0'). $\Delta \text{PA}$ indicates the angle between the VLBI jet and the optical minor axis of its host galaxy. The dark bands show the expected null signal (no alignment), which is the result of our careful analysis of the systematic errors. The stronger the deviation from the dark band the stronger the alignment signal. $p_5$ and $p_2$ are the p-values for five bins and two bins in the $\Delta\text{PA}$ space respectively. $p_{\text{ks}}$ is the p-value from a KS test comparing the data against a uniform distribution, independently of the number of bins. Each panel corresponds to a specific selection of sources, showing the number of sources on top. Each panel corresponds to a specific selection of sources based on several cuts. The number of sources inside each panel is shown on top. a) All `criterion 0' sources. b) All `criterion 0' sources whose jet and optical PA error is $<$ $22.5^{\circ}$. c) All `criterion 0' sources within the `good case'. d) All `criterion 0' sources with a measured spectroscopic redshift. e) All `criterion 0' sources whose jet and optical PA error is $<$ $22.5^{\circ}$ and with a measured spectroscopic redshift. f) All `criterion 0' sources within the `good case' and with a measured spectroscopic redshift. g) All `criterion 0' sources with $z < 0.5$. h) All `criterion 0' sources whose jet and optical PA error is $<$ $22.5^{\circ}$ and with $z < 0.5$. i) All `criterion 0' sources within the `good case' and with $z < 0.5$. j) All `criterion 0' sources with $z < 0.1$ k) All `criterion 0' sources whose jet and optical PA error is $<$ $22.5^{\circ}$ and with $z < 0.1$. l) All `criterion 0' sources within the `good case' and with $z < 0.1$.}
    \label{fig:hist_SM}
\end{figure*}

\begin{figure*}[h!]
    \includegraphics[width=\textwidth]{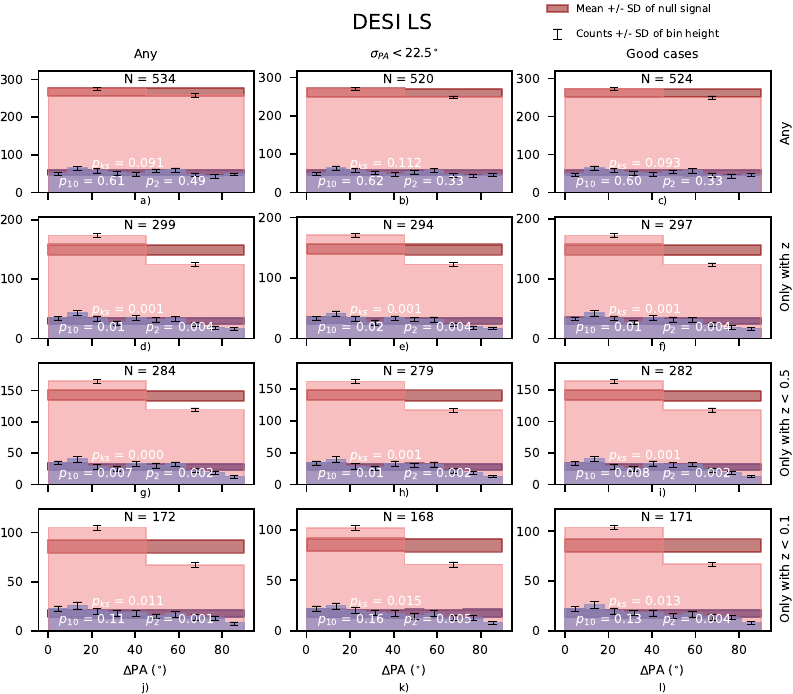}
    \caption{All VLBI sources with a cross-match in DESI LS and a semi-minor axis $>$ 1.3" (the `criterion 0'). $\Delta \text{PA}$ indicates the angle between the VLBI jet and the optical minor axis of its host galaxy. The dark bands show the expected null signal (no alignment), which is the result of our careful analysis of the systematic errors. The stronger the deviation from the dark band the stronger the alignment signal. $p_{10}$ and $p_2$ are the p-values for ten bins and two bins in the $\Delta\text{PA}$ space respectively. $p_{\text{ks}}$ is the p-value from a KS test comparing the data against a uniform distribution, independently of the number of bins. Each panel corresponds to a specific selection of sources, showing the number of sources on top. Each panel corresponds to a specific selection of sources based on several cuts. The number of sources inside each panel is shown on top. a) All `criterion 0' sources. b) All `criterion 0' sources whose jet and optical PA error is $<$ $22.5^{\circ}$. c) All `criterion 0' sources within the `good case'. d) All `criterion 0' sources with a measured spectroscopic redshift. e) All `criterion 0' sources whose jet and optical PA error is $<$ $22.5^{\circ}$ and with a measured spectroscopic redshift. f) All `criterion 0' sources within the `good case' and with a measured spectroscopic redshift. g) All `criterion 0' sources with $z < 0.5$. h) All `criterion 0' sources whose jet and optical PA error is $<$ $22.5^{\circ}$ and with $z < 0.5$. i) All `criterion 0' sources within the `good case' and with $z < 0.5$. j) All `criterion 0' sources with $z < 0.1$ k) All `criterion 0' sources whose jet and optical PA error is $<$ $22.5^{\circ}$ and with $z < 0.1$. l) All `criterion 0' sources within the `good case' and with $z < 0.1$.}
    \label{fig:DESI_10_bins}
\end{figure*}

\begin{figure*}[h!]             
    \includegraphics[width=\textwidth]{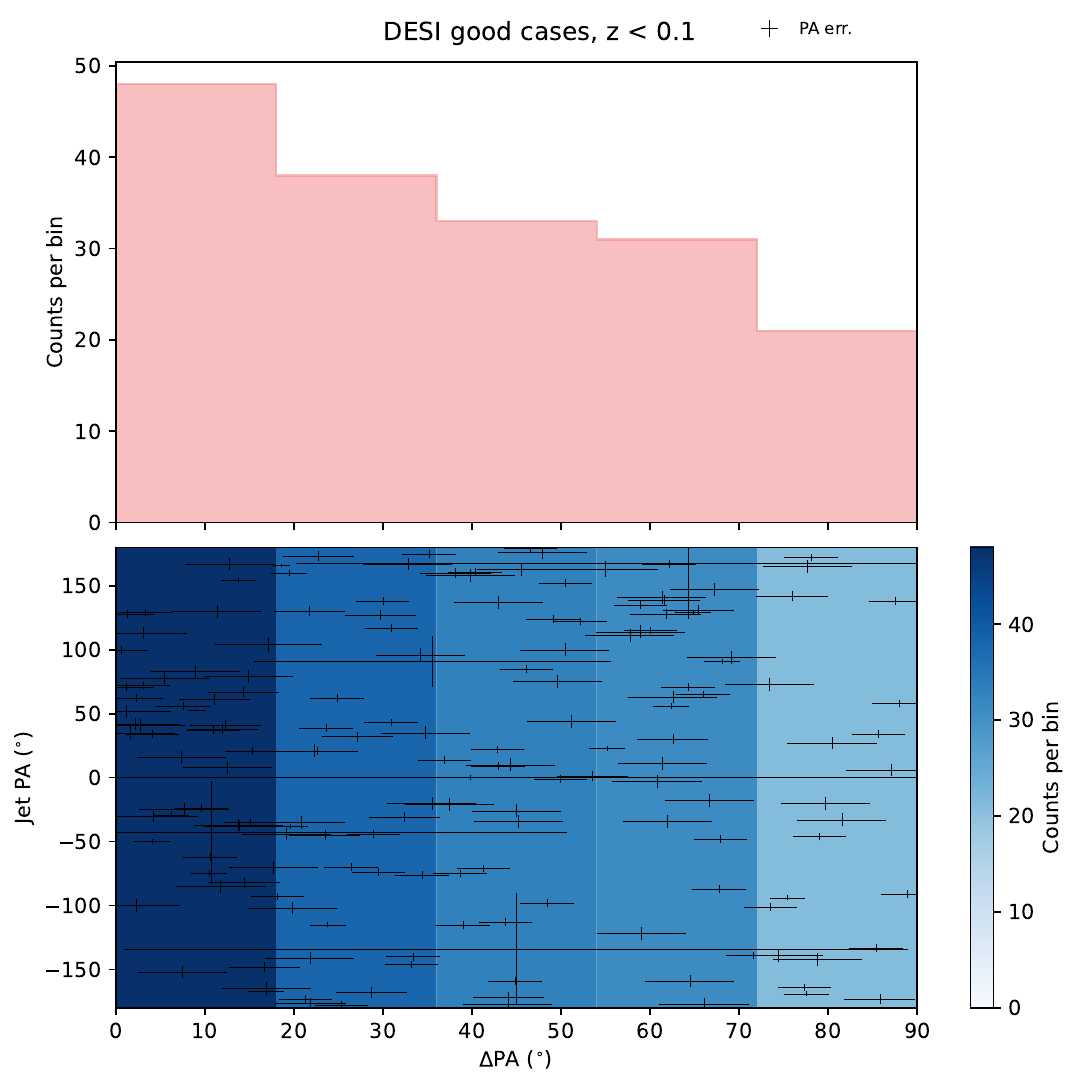}
    \caption{Scatter plot and histogram of VLBI sources with a cross-match in DESI LS, fulfilling the `criterion 0' (semi-minor axis $>$ 1.3"), the `good' case criteria and with z $<$ 0.1. The bottom panel shows the distribution of points in a scatter plot of $\Delta \text{PA}$ vs. Jet PA. The plot is divided into five vertical bands, each one colored according to the number of sources that fall within. The top panel shows a histogram with the number of sources in each band, recovering the same result as in the bottom right histogram of Fig. \ref{fig:hist_desi}.}
    \label{fig:DESI_scatter_hist}
\end{figure*}

\end{document}